\documentclass[floatfix,aps,prd,twocolumn,superscriptaddress,showpacs,amsmath,amssymb,letterpaper]{revtex4}

\usepackage{graphicx}
\usepackage{amsmath, amssymb}
\usepackage{dcolumn}
\usepackage{bm}
\usepackage{xspace}
\usepackage{multirow}
\usepackage{booktabs}
\usepackage{longtable}
\usepackage{textcomp}
\usepackage{subfigure}
\usepackage{lineno}

\usepackage{definitions}

\newcommand{\BaBar}{\mbox{\sl B\hspace{-0.4em} {\small\sl A}\hspace{-0.37em} \sl B\hspace{-0.4em} {\small\sl A\hspace{-0.02em}R}}}

\catcode`\@=11 
\newcommand{\parenbar}{\mathpalette\p@renb@r}
\def\p@renb@r#1#2{\vbox{%
  \ifx#1\scriptscriptstyle \dimen@.7em\dimen@ii.2em\else
  \ifx#1\scriptstyle \dimen@.8em\dimen@ii.25em\else
  \dimen@1em\dimen@ii.4em\fi\fi \offinterlineskip
  \ialign{\hfill##\hfill\cr
    \vbox{\hrule width\dimen@ii}\cr
    \noalign{\vskip-.3ex}%
    \hbox to\dimen@{$\mathchar300\hfil\mathchar301$}\cr
    \noalign{\vskip-.3ex}%
    $#1#2$\cr}}}
\catcode`\@=12 

\begin{document}


\title{{\boldmath Measurement of \textit{CP}-violation asymmetries in $D^0 \to K_S^0\,\pi^+\,\pi^-$}
}


\makeatletter
\twocolumn@sw{
\affiliation{Institute of Physics, Academia Sinica, Taipei, Taiwan 11529, Republic of China}
\affiliation{Argonne National Laboratory, Argonne, Illinois 60439, USA}
\affiliation{University of Athens, 157 71 Athens, Greece}
\affiliation{Institut de Fisica d'Altes Energies, ICREA, Universitat Autonoma de Barcelona, E-08193, Bellaterra (Barcelona), Spain}
\affiliation{Baylor University, Waco, Texas 76798, USA}
\affiliation{Istituto Nazionale di Fisica Nucleare Bologna, $^{ee}$University of Bologna, I-40127 Bologna, Italy}
\affiliation{University of California, Davis, Davis, California 95616, USA}
\affiliation{University of California, Los Angeles, Los Angeles, California 90024, USA}
\affiliation{Instituto de Fisica de Cantabria, CSIC-University of Cantabria, 39005 Santander, Spain}
\affiliation{Carnegie Mellon University, Pittsburgh, Pennsylvania 15213, USA}
\affiliation{Enrico Fermi Institute, University of Chicago, Chicago, Illinois 60637, USA}
\affiliation{Comenius University, 842 48 Bratislava, Slovakia; Institute of Experimental Physics, 040 01 Kosice, Slovakia}
\affiliation{Joint Institute for Nuclear Research, RU-141980 Dubna, Russia}
\affiliation{Duke University, Durham, North Carolina 27708, USA}
\affiliation{Fermi National Accelerator Laboratory, Batavia, Illinois 60510, USA}
\affiliation{University of Florida, Gainesville, Florida 32611, USA}
\affiliation{Laboratori Nazionali di Frascati, Istituto Nazionale di Fisica Nucleare, I-00044 Frascati, Italy}
\affiliation{University of Geneva, CH-1211 Geneva 4, Switzerland}
\affiliation{Glasgow University, Glasgow G12 8QQ, United Kingdom}
\affiliation{Harvard University, Cambridge, Massachusetts 02138, USA}
\affiliation{Division of High Energy Physics, Department of Physics, University of Helsinki and Helsinki Institute of Physics, FIN-00014, Helsinki, Finland}
\affiliation{University of Illinois, Urbana, Illinois 61801, USA}
\affiliation{The Johns Hopkins University, Baltimore, Maryland 21218, USA}
\affiliation{Institut f\"{u}r Experimentelle Kernphysik, Karlsruhe Institute of Technology, D-76131 Karlsruhe, Germany}
\affiliation{Center for High Energy Physics: Kyungpook National University, Daegu 702-701, Korea; Seoul National University, Seoul 151-742, Korea; Sungkyunkwan University, Suwon 440-746, Korea; Korea Institute of Science and Technology Information, Daejeon 305-806, Korea; Chonnam National University, Gwangju 500-757, Korea; Chonbuk National University, Jeonju 561-756, Korea}
\affiliation{Ernest Orlando Lawrence Berkeley National Laboratory, Berkeley, California 94720, USA}
\affiliation{University of Liverpool, Liverpool L69 7ZE, United Kingdom}
\affiliation{University College London, London WC1E 6BT, United Kingdom}
\affiliation{Centro de Investigaciones Energeticas Medioambientales y Tecnologicas, E-28040 Madrid, Spain}
\affiliation{Massachusetts Institute of Technology, Cambridge, Massachusetts 02139, USA}
\affiliation{Institute of Particle Physics: McGill University, Montr\'{e}al, Qu\'{e}bec, Canada H3A~2T8; Simon Fraser University, Burnaby, British Columbia, Canada V5A~1S6; University of Toronto, Toronto, Ontario, Canada M5S~1A7; and TRIUMF, Vancouver, British Columbia, Canada V6T~2A3}
\affiliation{University of Michigan, Ann Arbor, Michigan 48109, USA}
\affiliation{Michigan State University, East Lansing, Michigan 48824, USA}
\affiliation{Institution for Theoretical and Experimental Physics, ITEP, Moscow 117259, Russia}
\affiliation{University of New Mexico, Albuquerque, New Mexico 87131, USA}
\affiliation{The Ohio State University, Columbus, Ohio 43210, USA}
\affiliation{Okayama University, Okayama 700-8530, Japan}
\affiliation{Osaka City University, Osaka 588, Japan}
\affiliation{University of Oxford, Oxford OX1 3RH, United Kingdom}
\affiliation{Istituto Nazionale di Fisica Nucleare, Sezione di Padova-Trento, $^{ff}$University of Padova, I-35131 Padova, Italy}
\affiliation{University of Pennsylvania, Philadelphia, Pennsylvania 19104, USA}
\affiliation{Istituto Nazionale di Fisica Nucleare Pisa, $^{gg}$University of Pisa, $^{hh}$University of Siena and $^{ii}$Scuola Normale Superiore, I-56127 Pisa, Italy}
\affiliation{University of Pittsburgh, Pittsburgh, Pennsylvania 15260, USA}
\affiliation{Purdue University, West Lafayette, Indiana 47907, USA}
\affiliation{University of Rochester, Rochester, New York 14627, USA}
\affiliation{The Rockefeller University, New York, New York 10065, USA}
\affiliation{Istituto Nazionale di Fisica Nucleare, Sezione di Roma 1, $^{jj}$Sapienza Universit\`{a} di Roma, I-00185 Roma, Italy}
\affiliation{Rutgers University, Piscataway, New Jersey 08855, USA}
\affiliation{Texas A\&M University, College Station, Texas 77843, USA}
\affiliation{Istituto Nazionale di Fisica Nucleare Trieste/Udine, I-34100 Trieste, $^{kk}$University of Udine, I-33100 Udine, Italy}
\affiliation{University of Tsukuba, Tsukuba, Ibaraki 305, Japan}
\affiliation{Tufts University, Medford, Massachusetts 02155, USA}
\affiliation{University of Virginia, Charlottesville, Virginia 22906, USA}
\affiliation{Waseda University, Tokyo 169, Japan}
\affiliation{Wayne State University, Detroit, Michigan 48201, USA}
\affiliation{University of Wisconsin, Madison, Wisconsin 53706, USA}
\affiliation{Yale University, New Haven, Connecticut 06520, USA}

\author{T.~Aaltonen}
\affiliation{Division of High Energy Physics, Department of Physics, University of Helsinki and Helsinki Institute of Physics, FIN-00014, Helsinki, Finland}
\author{B.~\'{A}lvarez~Gonz\'{a}lez$^z$}
\affiliation{Instituto de Fisica de Cantabria, CSIC-University of Cantabria, 39005 Santander, Spain}
\author{S.~Amerio}
\affiliation{Istituto Nazionale di Fisica Nucleare, Sezione di Padova-Trento, $^{ff}$University of Padova, I-35131 Padova, Italy}
\author{D.~Amidei}
\affiliation{University of Michigan, Ann Arbor, Michigan 48109, USA}
\author{A.~Anastassov$^x$}
\affiliation{Fermi National Accelerator Laboratory, Batavia, Illinois 60510, USA}
\author{A.~Annovi}
\affiliation{Laboratori Nazionali di Frascati, Istituto Nazionale di Fisica Nucleare, I-00044 Frascati, Italy}
\author{J.~Antos}
\affiliation{Comenius University, 842 48 Bratislava, Slovakia; Institute of Experimental Physics, 040 01 Kosice, Slovakia}
\author{G.~Apollinari}
\affiliation{Fermi National Accelerator Laboratory, Batavia, Illinois 60510, USA}
\author{J.A.~Appel}
\affiliation{Fermi National Accelerator Laboratory, Batavia, Illinois 60510, USA}
\author{T.~Arisawa}
\affiliation{Waseda University, Tokyo 169, Japan}
% ...

\collaboration{CDF Collaboration\footnote{With visitors from
$^a$Istituto Nazionale di Fisica Nucleare, Sezione di Cagliari, 09042 Monserrato (Cagliari), Italy,
$^b$University of CA Irvine, Irvine, CA 92697, USA,
$^c$University of CA Santa Barbara, Santa Barbara, CA 93106, USA,
$^d$University of CA Santa Cruz, Santa Cruz, CA 95064, USA,
$^e$Institute of Physics, Academy of Sciences of the Czech Republic, Czech Republic,
$^f$CERN, CH-1211 Geneva, Switzerland,
$^g$Cornell University, Ithaca, NY 14853, USA,
$^h$University of Cyprus, Nicosia CY-1678, Cyprus,
$^i$Office of Science, U.S. Department of Energy, Washington, DC 20585, USA,
$^j$University College Dublin, Dublin 4, Ireland,
$^k$ETH, 8092 Zurich, Switzerland,
$^l$University of Fukui, Fukui City, Fukui Prefecture, Japan 910-0017,
$^m$Universidad Iberoamericana, Mexico D.F., Mexico,
$^n$University of Iowa, Iowa City, IA 52242, USA,
$^o$Kinki University, Higashi-Osaka City, Japan 577-8502,
$^p$Kansas State University, Manhattan, KS 66506, USA,
$^q$Ewha Womans University, Seoul, 120-750, Korea,
$^r$University of Manchester, Manchester M13 9PL, United Kingdom,
$^s$Queen Mary, University of London, London, E1 4NS, United Kingdom,
$^t$University of Melbourne, Victoria 3010, Australia,
$^u$Muons, Inc., Batavia, IL 60510, USA,
$^v$Nagasaki Institute of Applied Science, Nagasaki, Japan,
$^w$National Research Nuclear University, Moscow, Russia,
$^x$Northwestern University, Evanston, IL 60208, USA,
$^y$University of Notre Dame, Notre Dame, IN 46556, USA,
$^z$Universidad de Oviedo, E-33007 Oviedo, Spain,
$^{aa}$CNRS-IN2P3, Paris, F-75205 France,
$^{bb}$Texas Tech University, Lubbock, TX 79609, USA,
$^{cc}$Universidad Tecnica Federico Santa Maria, 110v Valparaiso, Chile,
$^{dd}$Yarmouk University, Irbid 211-63, Jordan,
$^{mm}$University of Warwick, Coventry CV4 7AL, United Kingdom
}}
\noaffiliation
}
\makeatother


\begin{abstract}
  We report a measurement of time-integrated \textit{CP}-violation
  asymmetries in the resonant substructure of the three-body decay
  $D^0 \to K_S^0\,\pi^+\,\pi^-$ using CDF II data corresponding to
  6.0\,\invfb of integrated luminosity from Tevatron $p\bar{p}$
  collisions at $\sqrt{s}$ = 1.96\,TeV. 
  The charm mesons used in this analysis come from $D^{*+}(2010)\rightarrow D^0\pi^+$ and
$D^{*-}(2010)\rightarrow \bar{D}^0\pi^-$, where the production flavor of the charm meson is
determined by the charge of the accompanying pion.
  We apply
  a Dalitz-amplitude analysis for the description of the dynamic decay
  structure and use two complementary approaches, namely a full
  Dalitz-plot fit employing the isobar model for the contributing
  resonances and  a model-independent bin-by-bin comparison of the
  $D^0$ and $\bar{D}^0$ Dalitz plots. We find no \textit{CP}-violation
  effects and measure an asymmetry of $A_{CP} = (-0.05 \pm 0.57
  (\mathrm{stat}) \pm 0.54 (\mathrm{syst}))\%$ for the overall
  integrated \textit{CP}-violation asymmetry, consistent with the
  standard model prediction.
\end{abstract}


\maketitle


\section{Introduction}
\label{sec:Intro}

The phenomenon of \textit{CP} violation is well established for weakly
decaying hadrons consisting of down-type quarks. 
It has been observed with mixing (indirect \textit{CP} violation) and without mixing (direct \textit{CP} violation) in
decays of the $K^0$ and $B^0$ mesons, and confirms the theory of Kobayashi and
Maskawa~\cite{Kobayashi:1973fv} that describes \textit{CP} violation
in the standard model. For charm mesons, \textit{CP}-violation effects
are expected to be small. 
The lack of experimental observation of \textit{CP} violation in charm meson decays is consistent with the
expectation.
Only recently the LHCb collaboration has reported the first
evidence for \textit{CP} violation in the time-integrated rate of
$D^0$ decays to two hadrons at
$\mathcal{O}(10^{-2})$~\cite{Aaij:2011in}, which was consequently confirmed by the CDF collaboration~\cite{pubnote:cdf10784}. 
Whether this is consistent
with the standard model expectation or a hint for new physics is not
yet clear. Therefore, it is important to complement the LHCb result
with measurements of \textit{CP} asymmetries in other $D^0$ decay
modes.

In this article, we describe a search for \textit{CP} violation
in time-integrated decay rates of $D^0$ mesons to the
$K_S^0\,\pi^+\,\pi^-$ final state. The standard model
expectations for the size of time-integrated \textit{CP}-violation
asymmetries in $K_S^0\,\pi^+\,\pi^-$ decays are
$\mathcal{O}(10^{-6})$, where the dominant contribution arises from
\textit{CP} violation in $K^0$-$\bar{K^0}$ mixing~\cite{Bigi:1994aw,D0KSPiPiCPV}. The
observation of any significantly larger asymmetries would be a strong
hint for physics beyond the standard model.

The CLEO collaboration performed a dedicated search for
time-integrated \textit{CP} violation in a fit to the Dalitz plot of
the $D^0/\bar{D}^0$ three-body decay to
$K_S^0\,\pi^+\,\pi^-$~\cite{CleoCPV}. Belle and $\BaBar$ allowed for
\textit{CP} violation in their measurements of the $D^0$-$\bar{D}^0$
mixing parameters~\cite{BelleMixing,BabarMixing}. Up to now, no
\textit{CP}-violation effects have been found~\cite{HFAG}.

In this analysis, we exploit a large sample of $D^*(2010)^{\pm}$
decays, reconstructed in a dataset corresponding to 6.0\,\invfb of integrated luminosity produced
in $p\bar{p}$ collisions at $\sqrt{s}$ = 1.96\,TeV and collected by
the CDF II detector, to measure time-integrated \textit{CP}-violation
asymmetries in the resonant substructure of the decay $D^0/\bar{D}^0
\to K_S^0\,\pi^+\,\pi^-$. The neutral $D$ meson production flavor is
determined by the charge of the pion in the 
$D^{*+}(2010)\rightarrow D^0\pi^+$ and
$D^{*-}(2010)\rightarrow \bar{D}^0\pi^-$ decays ($D^*$ tagging).
Throughout the rest
of the paper the use of a specific particle state implies the use of
the charge-conjugate state as well, unless explicitly noted. For
brevity, $D^*(2010)^+$ is abbreviated as $D^{*+}$.

In Sections~\ref{sec:detector} and \ref{sec:trigger} we briefly
describe the CDF II detector and the trigger components important for
this analysis. In Sections~\ref{sec:reconstruction} and
\ref{sec:selection} we describe the offline candidate reconstruction
and selection, respectively. In Sec.~\ref{sec:fit} we explain the
Dalitz-plot fits of the resonant substructure of the decay, followed
by a discussion of systematic uncertainties in
Sec.~\ref{sec:systematics} and a presentation of the results in
Sec.~\ref{sec:results}. In Sec.~\ref{sec:CPV_Binned} we describe a
model-independent search for \textit{CP} asymmetries using a
bin-by-bin comparison of Dalitz plots~\cite{Bigi}, followed by the
conclusion in Sec.~\ref{sec:conclusion}.

\section{CDF II detector}
\label{sec:detector}

The analysis is performed on a dataset collected with the CDF II
detector~\cite{Acosta:2004yw} between February 2002 and February 2010,
corresponding to an integrated luminosity of $6.0\,\mathrm{fb^{-1}}$
of $p\bar{p}$ collisions. Among the components and capabilities of the
detector, the charged particle tracking is the one most relevant to
this analysis. The tracking system lies within a uniform, axial
magnetic field of $1.4$\,T. The inner tracking volume, up to a radius
of $28$\,cm, is composed of six or seven layers, depending on polar
angle, of double-sided silicon microstrip detectors~\cite{SVX-II,ISL}.
An additional single-sided silicon layer is mounted directly on the
beam pipe at a radius of $1.5$\,cm~\cite{Hill:2004qb}, allowing
excellent resolution of the transverse impact parameter $d_0$, defined
as the distance of closest approach of a charged particle trajectory
(track) to the interaction point in the plane transverse to the beam
line. The silicon detector allows the identification of displaced,
secondary vertices from bottom-hadron and charm-hadron decays with a
resolution of approximately $30\,\text{\textmu} \mathrm{m}$ in the
transverse and $70\,\text{\textmu} \mathrm{m}$ in the longitudinal
direction. The outer tracking volume from a radius of $40$ to
$137$\,cm is occupied by an open-cell argon-ethane gas drift chamber
(COT)~\cite{Affolder:2003ep}. An important aspect for this analysis is
that the layout of the COT is intrinsically charge-asymmetric because
of an about $35^\circ$ tilt angle between the cell orientation and the
radial direction. The total tracking system provides a transverse
momentum resolution of $\sigma(p_T)/p_T^2 \approx
0.07\%\,(\gevc)^{-1}$ for tracks with $p_T > 2\,\gevc$. A more
detailed description of the tracking system can be found in
Ref.~\cite{D0hh}.

\section{Online event selection}
\label{sec:trigger}

A three-level  event-selection system (trigger) is used. At
level 1, a hardware track-processor~\cite{Thomson:2002xp} identifies
charged particles using information from the COT and measures their
transverse momenta and azimuthal angles around the beam direction. The
basic requirement at level 1 is the presence of two charged particles,
each with $p_T > 2\,\gevc$. At level 2, the silicon vertex
trigger~\cite{Ristori:2010} adds silicon-hit information to the
tracks found by the hardware track-processor, thus allowing the precise measurement of impact parameters of
tracks. The two level-1 tracks are required to have impact parameters
between $0.1$ and 1\,mm, an opening angle in the transverse plane between 2$^\circ$ and
90$^\circ$, and to be consistent with coming from a common
vertex displaced from the interaction point by at least
200\,\text{\textmu}m in the plane transverse to the beam line. 
This is complemented by a selection without the vertex displacement requirement, collecting events with low invariant-mass track pairs having opening angles less than 6$^\circ$. 
The level-3 trigger is implemented in software and provides the final
online selection by confirming the first two trigger-level decisions
using a more precise reconstruction. This trigger is designed to
collect hadronic decays of long-lived particles such as $b$ and $c$
hadrons. Three different configurations of this trigger are employed,
requiring a minimum on the scalar sum of the transverse momenta of the
two trigger tracks of 4.0, 5.5, and 6.5\,\gevc. The active threshold
depends on the instantaneous luminosity conditions, with higher
thresholds used at higher instantaneous luminosity to reduce the
higher trigger accept-rate.

\section{Offline event reconstruction}
\label{sec:reconstruction}

The offline reconstruction of candidates starts by fitting tracks
taking into account multiple scattering and ionization energy-loss
calculated for the pion mass hypothesis. 
Since all final state particles in the studied decay chain are pions, we assign the pion mass to each
track in the following steps.
Two oppositely-charged tracks are combined to form a $K_S^0$ candidate. To
construct $D^0$ candidates, each $K_S^0$ candidate is  combined
with all other possible oppositely charged track pairs found in the
event. Finally, the $D^{*+}$ candidates are obtained by combining each
$D^0$ candidate with one of the remaining tracks in the event. The
tracks forming the $K_S^0$, $D^0$, and $D^{*+}$ candidates are
subjected to separate kinematic fits that constrain them to originate
from a common decay point in each case, resulting in three reconstructed decay points for the
considered decay chain. In each step of the reconstruction, standard
quality requirements on tracks and vertices are used to ensure
well-measured momenta and decay positions~\cite{Wick:2011zz}.

\section{Candidate selection}
\label{sec:selection}

For the selection of the candidates, we first impose some quality
requirements to suppress the most obvious backgrounds, such as random
combinations of low-$p_T$ particles, $D^{*+}$ pions strongly displaced
from the interaction point, or short-lived $D^0$ candidates without
proper secondary-vertex separation. We require the transverse momentum
of each pion to be greater than 400\,\mevc, the transverse momentum of
the $D^{*+}$ to exceed 5.0\,\gevc, the impact parameter of the pion
from the $D^{*+}$ decay divided by its uncertainty
$d_0/\sigma_{d_0}(\pi_{D^{*+}})$ not to exceed 15, and the transverse
decay length of the $D^0$ candidate projected into the transverse
momentum ($L_{xy}$) to exceed its resolution ($\sigma_{L_{xy}}$). For
the surviving candidates we use an artificial neural network to
distinguish signal from background. The neural network is constructed
with the NeuroBayes package~\cite{Feindt:2006pm} and trained, using
data only, by means of the $_s\mathcal{P}lot$
technique~\cite{Pivk:2004ty}. This technique assigns a weight to each
candidate, proportional to the probability that the candidate is
signal. The candidate weight is based on a discriminating variable,
which is required to be independent of the ones used in the neural
network training. In our case, the discriminating variable is the mass
difference $\Delta M = M(K_S^0\pi^+\pi^-\,\pi^+) - M(K_S^0\pi^+\pi^-)$
of the $D^{*+}$ candidate. In the training, each candidate enters with
a weight calculated from the signal probability that is derived from
its mass. Based on these weights, the neural network learns the
features of signal and background events. Since we use only data for
the neural network training, we randomly split each sample into two
parts evenly distributed in data taking time and train two networks.
Each network is then applied to the complementary subsample in order
for the selection to be trained on a sample independent from the one
to which it is applied. This approach avoids a bias of the selection
originating from statistical fluctuations possibly learned by the
network. The method of NeuroBayes $_s\mathcal{P}lot$ trainings was
first applied in our previous work on charm baryons~\cite{ourPRD}.

The network uses five input variables. Ordered by decreasing
importance, these are $L_{xy}/\sigma_{L_{xy}}(D^0)$, the $\chi^2$
quality of the $D^{*+}$ vertex fit, $d_0/\sigma_{d_0}(\pi_{D^{*+}})$,
$p_T(\pi_{D^{*+}})$, and the reconstructed mass of the $K_S^0$
candidate. The $D^{*+}$ network training is based on the mass
difference distribution in the range $140 < \Delta M < 156\,\mevcc$. A
fit of a non-relativistic Breit-Wigner convoluted with a Gaussian for
the signal and a third-order polynomial for the background function
defines the probability density functions used to calculate the
$_s\mathcal{P}lot$ weights. The final neural network output
requirement is chosen to maximize $S/\sqrt{S+B}$ where $S$ ($B$) is
the estimated number of signal (background) events in the signal
region estimated from a fit to the $M(K_S^0\pi^+\pi^-)$ distribution.
The selected requirement corresponds to an \textit{a posteriori}
signal probability greater than $25\%$. In 11\% of the selected
events, multiple $D^{*+}$ candidates are reconstructed for a single
$D^0$ candidate. We choose the $D^{*+}$ candidate that gives the most
$D^*$-like neural network output and remove all others to avoid
identical candidates populating the Dalitz space.

\begin{figure}
  \includegraphics[width=8.5cm]{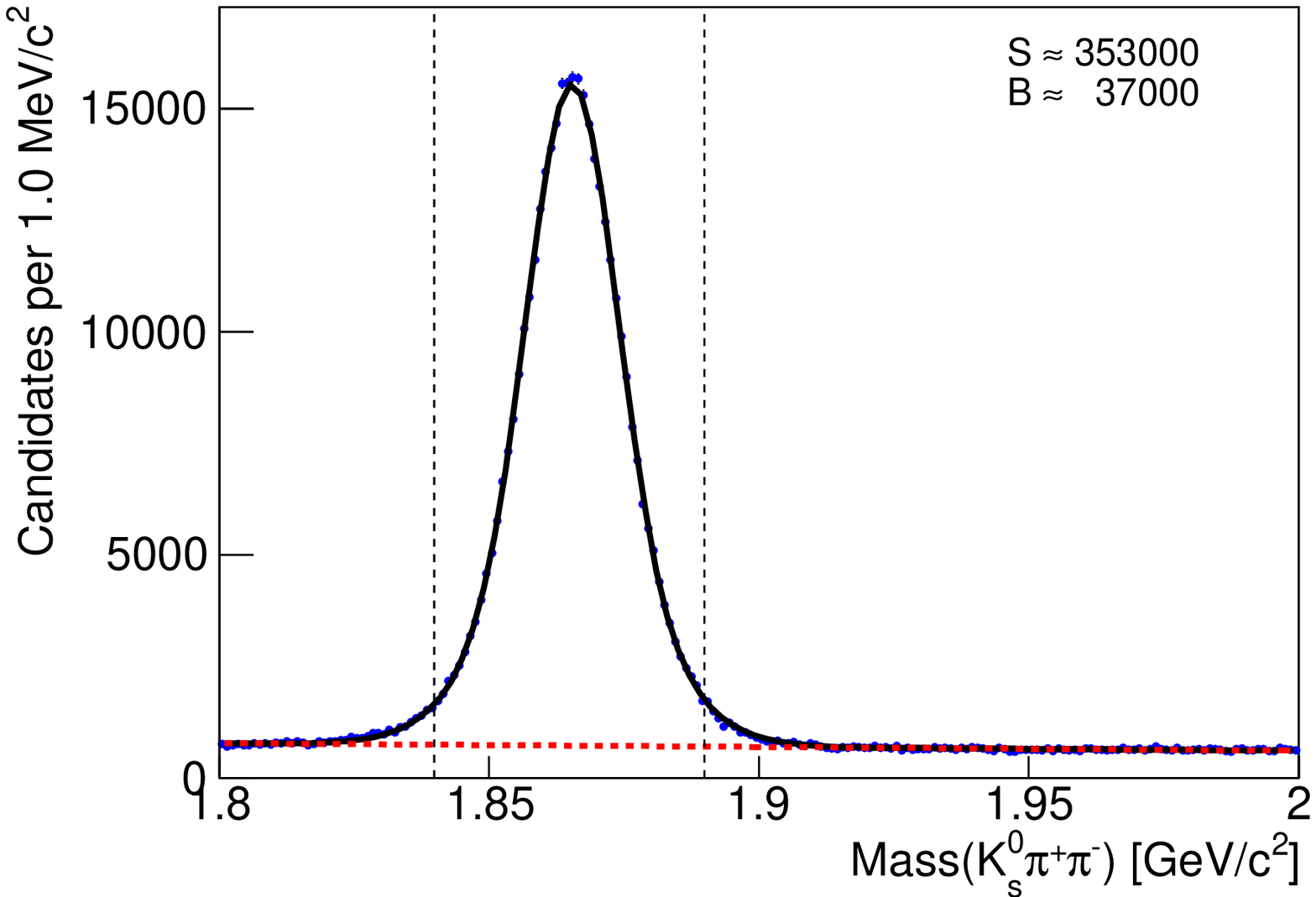}
  \includegraphics[width=8.5cm]{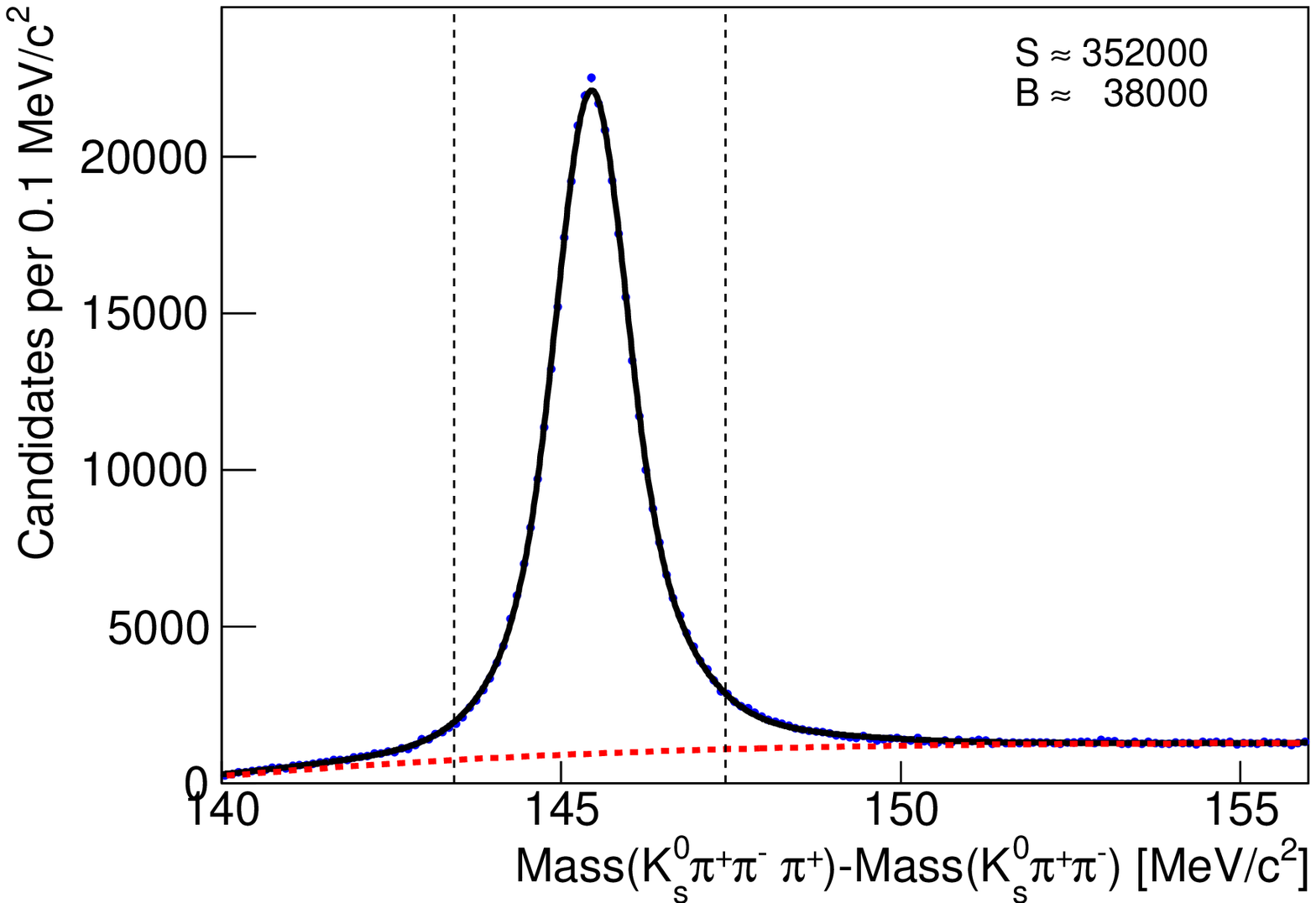}
  \caption{The $M(K_S^0\pi^+\pi^-)$ and $\Delta M$ data distributions
    (points with error bars) with fit results overlaid. The dashed
    lines correspond to the background contributions. The vertical
    lines indicate signal ranges used for further analysis. The
    $M(K_S^0\pi^+\pi^-)$ ($\Delta M$) distribution contains only
    candidates populating the signal $\Delta M$ ($M(K_S^0\pi^+\pi^-)$)
    range.}
  \label{fig:MassFits}
\end{figure}

The $M(K_S^0\pi^+\pi^-)$ and $\Delta M$ distributions of the selected
candidates are shown in Fig.~\ref{fig:MassFits}, together with the
corresponding fits to determine the signal and background yields. In
the $K_S^0\pi^+\pi^-$ mass distribution, the $D^0$ signal is described
by the sum of two Gaussian functions with common mean, and the
background is modeled by a linear function. The $D^{*+}$ signal in the
$\Delta M$ distribution is described by a non-relativistic
Breit-Wigner convoluted with a resolution function. The latter is
determined from simulated events and consists of the weighted sum of
three Gaussian functions to model the more complicated shape of the
resolution close to the kinematic threshold in the $\Delta M$
distribution. The background is modeled by a third-order polynomial.

For the Dalitz-plot studies, the analysis is restricted to candidates
populating two mass ranges, $1.84 < M(K_S^0\pi^+\pi^-) < 1.89\,\gevcc$
and $143.4 < \Delta M < 147.4\,\mevcc$, indicated by the dashed
vertical lines in Fig.~\ref{fig:MassFits}. The selected data sample
contains approximately $3.5 \times 10^5$ signal events and consists of
about 90\% correctly $D^*$-tagged $D^0$ signal, 1\% mistagged $D^0$
signal, and 9\% background candidates. A mistag results from the
combination of a $D^0$ with a random, wrongly-charged pion not
originating from a $D^{*\pm}$ decay. The mistag fraction is estimated
by a dedicated parameter in the Dalitz-plot fit described in
Sec.~\ref{sec:comb_fit}.

\begin{figure}
  \centerline{\includegraphics[width=8.5cm]{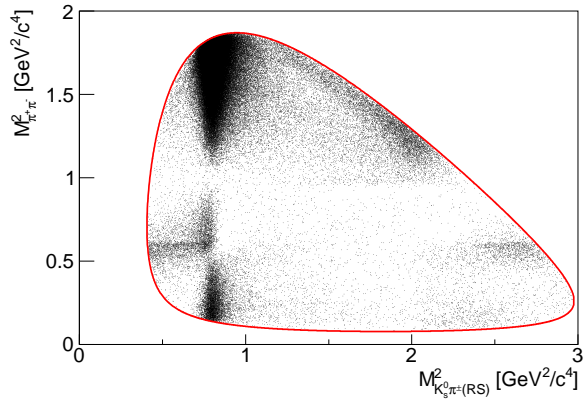}}
  \caption{Dalitz plot of the decay $D^0/\bar{D}^0 \to
    K_S^0\,\pi^+\,\pi^-$, where the squared invariant masses of the
    two-body combinations $K_S^0\pi^{\pm}(\mathrm{RS})$ and
    $\pi^+\pi^-$ are chosen as kinematic quantities. The solid line
    indicates the kinematic boundaries.}
  \label{fig:DalitzPlot}
\end{figure}

The resonant substructure of a three-body decay can be described by
means of a Dalitz-amplitude analysis~\cite{dalitz_orig}. The Dalitz
plot of the considered decay $D^0 \to K_S^0\,\pi^+\,\pi^-$, composed
of all selected candidates, is shown in Fig.~\ref{fig:DalitzPlot}. It
represents the decay dynamics as a function of the squared invariant
masses of the two-body combinations $K_S^0\pi^{\pm}(\mathrm{RS})$ and
$\pi^+\pi^-$, where the notation $K_S^0\pi^{\pm}(\mathrm{RS})$
expresses that the Cabibbo-favored combination, or right-sign (RS)
pion, is used for both $D^0$ ($K_S^0\pi^-$) and $\bar{D}^0$
($K_S^0\pi^+$) decays. The squared invariant mass of the
doubly-Cabibbo-suppressed or wrong-sign (WS) two-body combination
$K_S^0\pi^{\pm}(\mathrm{WS})$ is a linear function of
$M^2_{K_S^0\pi^{\pm}(\mathrm{RS})}$ and $M^2_{\pi^+\pi^-}$. Three
types of final states contribute in the decay, Cabibbo-favored,
doubly-Cabibbo-suppressed, and \textit{CP} eigenstates. The dominant
decay mode is the Cabibbo-favored $D^0 \to K^*(892)^-\,\pi^+$, which
amounts for about 60\% of the total branching fraction. The second
largest contribution is from the intermediate \textit{CP} eigenstate
$K_S^0\rho(770)$, which is color-suppressed compared to
$K^*(892)^-\pi^+$.

\section{Dalitz analysis}
\label{sec:fit}

A simultaneous fit to the resonant substructure of the decay to
$K_S^0\,\pi^+\,\pi^-$ is performed on the combined $D^0$ and
$\bar{D}^0$ samples to determine the sizes of the various
contributions. These are compared with previous results from other
experiments to build confidence in the fitting technique. Then the fit
is applied independently to $D^*$-tagged $D^0$ and $\bar{D}^0$ samples
to measure \textit{CP}-violation asymmetries in the decay amplitudes
for each subprocess.

\subsection{Fit to the combined sample}
\label{sec:comb_fit}

A binned maximum-likelihood fit to the two-dimensional Dalitz-plot
distribution with bin widths of $0.025\,\mathrm{GeV}^2/c^4$ in both
dimensions is performed to determine the contributions of the various
intermediate resonances. The likelihood function has the general form
\begin{eqnarray}
  \mathcal{L}(\vec{\theta}) = \prod_{i=1}^{I} \frac{\mu_{i}^{n_i} e^{- \mu_i}}{n_i!}\,,
  \label{eq:Likelihood}
\end{eqnarray}
where $\vec{\theta}$ are the estimated parameters, $I$ is the number
of bins, $n_i$ is the number of entries in bin $i$, and $\mu_i$ is the
expected number of entries in bin $i$. The latter are obtained using
the function
\begin{eqnarray}
  \mu(M_{K_S^0\pi^{\pm}(\mathrm{RS})}^2,M_{\pi^+\pi^-}^2) &=& N [T \epsilon(M_{K_S^0\pi^{\pm}(\mathrm{RS})}^2,M_{\pi^+\pi^-}^2) \nonumber \\
  && |\mathcal M(M_{K_S^0\pi^{\pm}(\mathrm{RS})}^2,M_{\pi^+\pi^-}^2)|^2 \nonumber \\
  &+& (1-T) \epsilon(M_{K_S^0\pi^{\pm}(\mathrm{RS})}^2,M_{\pi^+\pi^-}^2) \nonumber \\
  && |\mathcal M(M_{K_S^0\pi^{\pm}(\mathrm{WS})}^2,M_{\pi^+\pi^-}^2)|^2] \nonumber \\
  &+& B(M_{K_S^0\pi^{\pm}(\mathrm{RS})}^2,M_{\pi^+\pi^-}^2)\,,
    \label{eq:Dalitz_likelihood}
\end{eqnarray}
where $\mathcal M$ is the complex matrix element of the decay, $(1-T)$
is the fraction of $D^0$ candidates with wrongly-determined production
flavor, called the mistag fraction, $N$ is the normalization of the
number of signal events, $\epsilon(M_{K_S^0\pi^{\pm}(\mathrm{RS})}^2,M_{\pi^+\pi^-}^2)$
is the relative trigger and reconstruction efficiency over the Dalitz
plot, and $B(M_{K_S^0\pi^{\pm}(\mathrm{RS})}^2,M_{\pi^+\pi^-}^2)$ is the background
distribution. The function is evaluated at the bin center to calculate
the expectation for $\mu_i$.

The isobar model~\cite{Isobar} is used to describe the matrix element
$\mathcal M$. The various resonances are modeled by complex numbers
$a_j e^{i\delta_j}$, where $j$ refers to the $j$th isobar composed of
the amplitude $a_j$ and the phase $\delta_j$, multiplied with the
individual complex matrix element $\mathcal{A}_j(M,\Gamma)$, which
depends on the mass $M$ and decay width $\Gamma$ of the resonance. The
phase convention is the same as in the CLEO analysis~\cite{CleoCPV}
and described in Ref.~\cite{Wick:2011zz}. The complex numbers are
added as
\begin{equation}
  \mathcal{M} = a_0 e^{i\delta_0} + \sum_j a_j e^{i\delta_j} \mathcal{A}_j(M,\Gamma)\,,
  \label{eq:Isobar}  
\end{equation}
where $a_0 e^{i\delta_0}$ represents a possible non-resonant
contribution. Since $a_j$ and $\delta_j$ are relative amplitudes and
phases, one resonance can be chosen as the reference. The amplitude
and phase of the $\rho(770)$, being the largest color-suppressed mode,
are fixed to the values $a_{\rho(770)}=1$ and $\delta_{\rho(770)}=0$,
respectively. The individual matrix elements,
$\mathcal{A}_j(M,\Gamma)$, correspond to normalized Breit-Wigner
shapes with Blatt-Weisskopf form factors~\cite{BlattWeisskopf}. A more
detailed description can be found in Ref.~\cite{dalitz_method}. For
the intermediate resonances $\rho(770)$ and $\rho(1450)$ decaying to
$\pi^+\pi^-$, the Breit-Wigner shape is replaced by the
Gounaris-Sakurai description~\cite{sakurai}.

To account for the limited accuracy of the knowledge on the masses and
widths of the intermediate resonances, these parameters can vary
within their experimental uncertainties, taken from
Ref.~\cite{f0masses} for the $f_0(980)$ and $f_0(1370)$ mesons, and
Ref.~\cite{PDG} for the others. This is accomplished by means of
Gaussian constraints in the likelihood function, except for the
resonances $K^*(892)^{\pm}$, $f_0(600)$, and $\sigma_2$, which are
unconstrained. Because the $K^*(892)^{\pm}$ is the most prominent
resonance, with its 60\% branching fraction, floating the
$K^*(892)^{\pm}$ mass and width in the Dalitz-plot fit improves the
fit quality. The reason for the unconstrained $f_0(600)$ and
$\sigma_2$ resonance parameters is the poorly known nature of these
states. The scalar resonance $\sigma_2$ is introduced to account for a
structure near $1\,\mathrm{GeV^2}/c^4$ in the $M^2_{\pi^+\pi^-}$
distribution. A possible explanation for this structure, proposed in
Ref.~\cite{Belle_gamma_old}, is the decay $f_0(980) \to \eta\,\eta$
with rescattering of $\eta\eta$ to $\pi^+\pi^-$, resulting in a
distortion of the $f_0(980) \to \pi^+\,\pi^-$ amplitude for
$M^2_{\pi^+\pi^-}$ near the $\eta\eta$ production threshold. The
masses and widths of the resonances $K^*(892)^{\pm}$,
$K_0^*(1430)^{\pm}$, and $K_2^*(1430)^{\pm}$ are required to be
identical for Cabibbo-favored and doubly-Cabibbo-suppressed (DCS)
processes.

\begin{figure}
  \includegraphics[width=8.5cm]{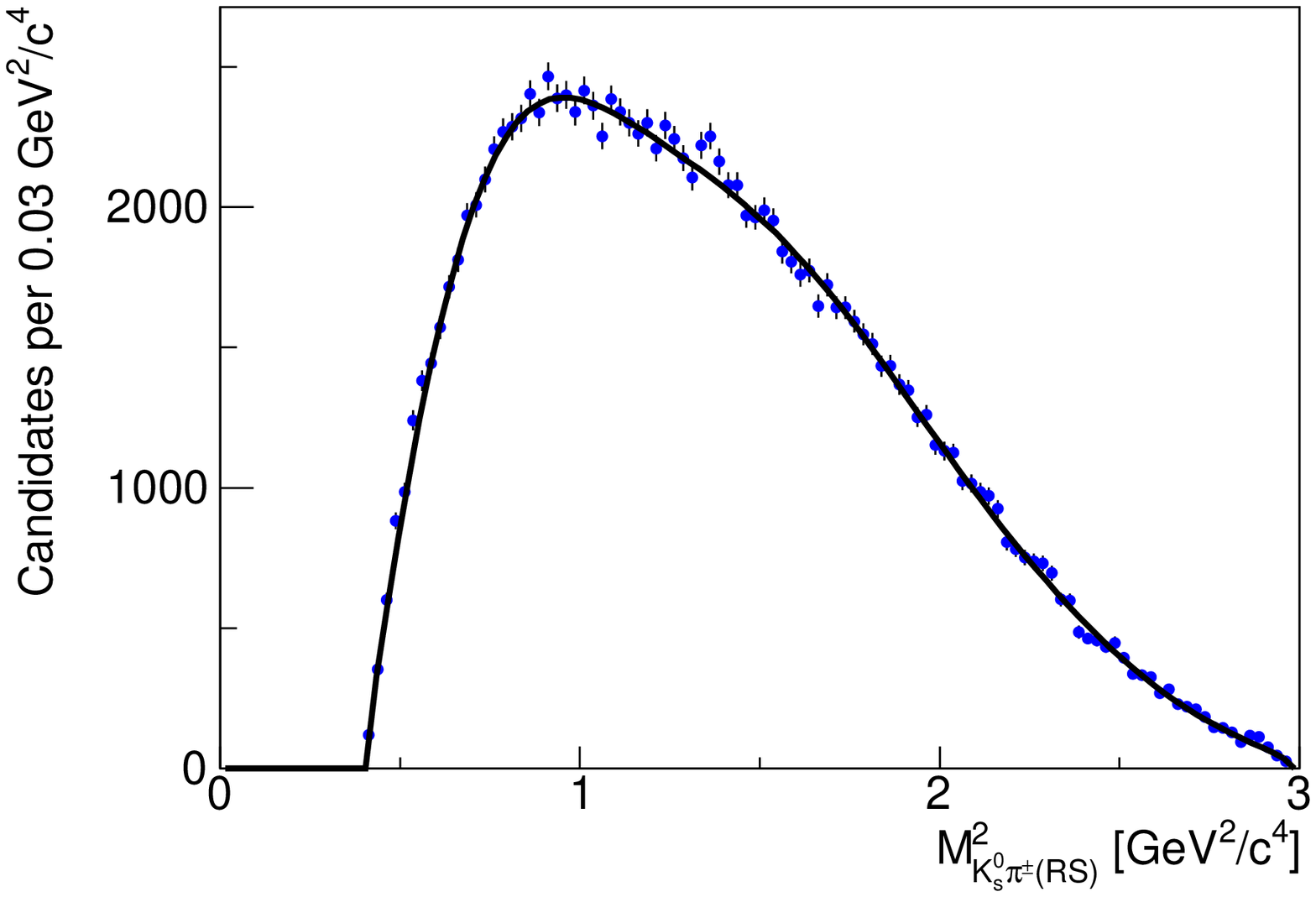}
  \includegraphics[width=8.5cm]{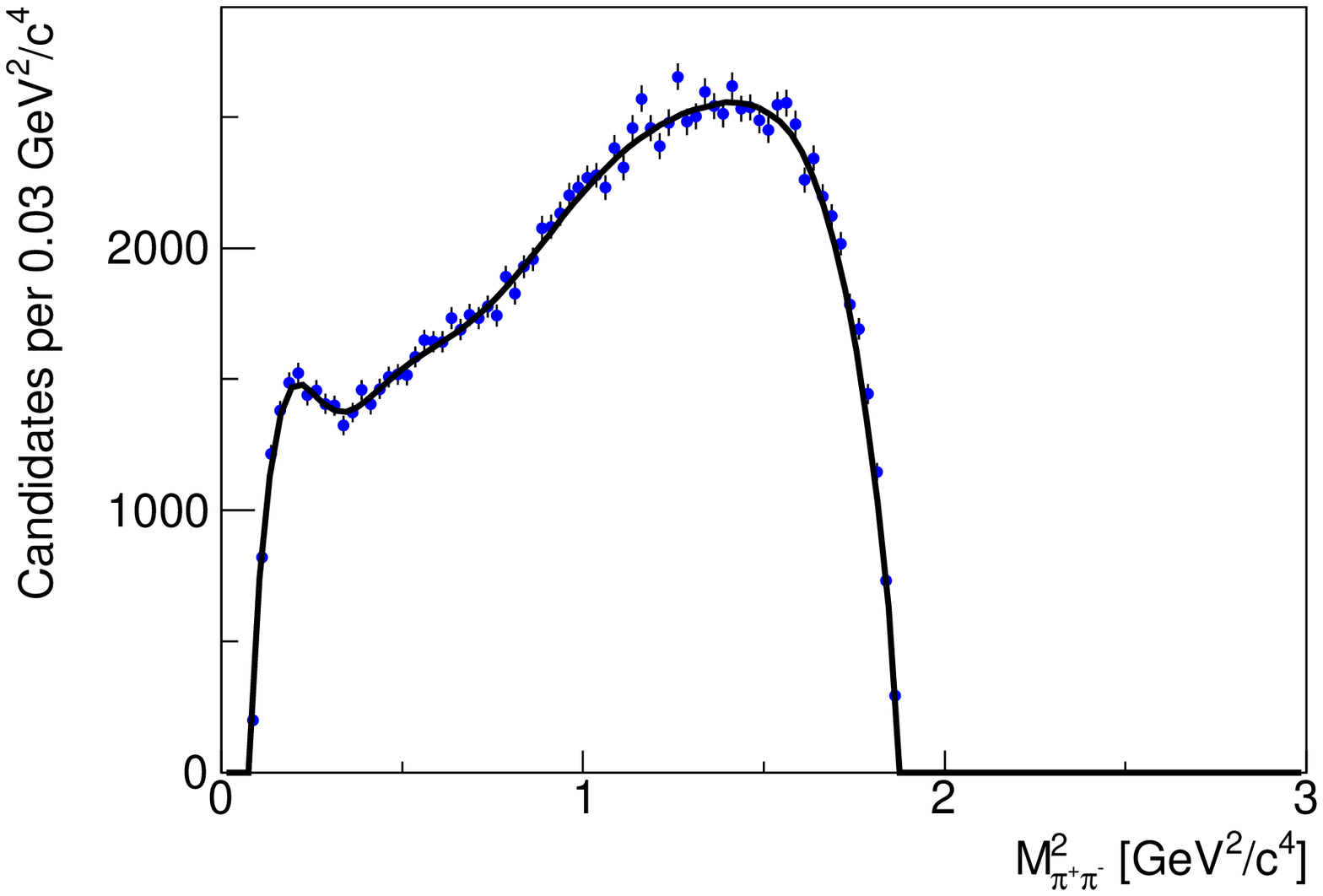}
  \includegraphics[width=8.5cm]{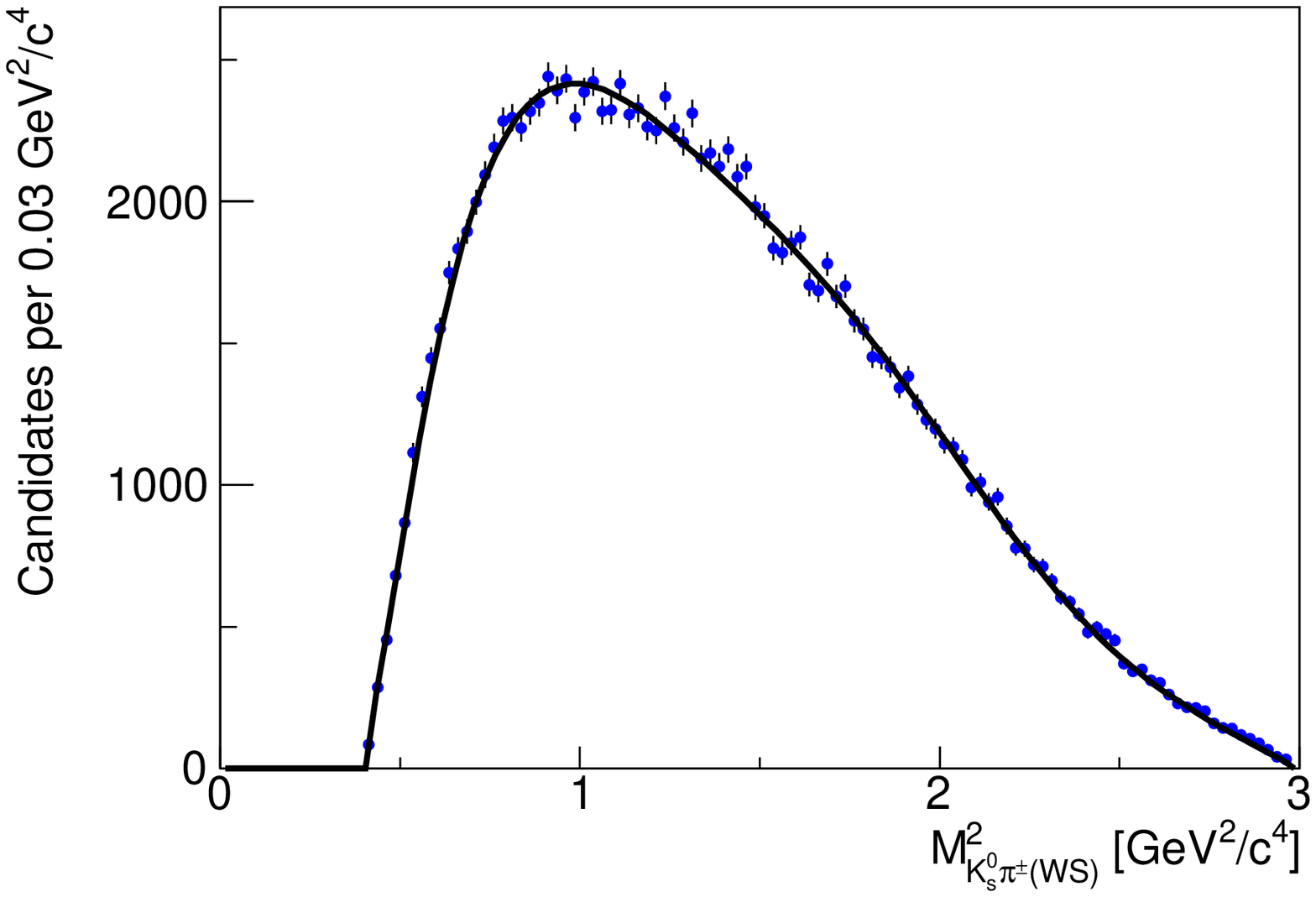}
  \caption{Squared invariant mass distributions of the simulated
    events used for the determination of the relative reconstruction
    efficiency over the Dalitz plot, together with the corresponding
    fit projections.}
  \label{fig:EfficiencyFitProj}
\end{figure}

Simulated events are used to estimate the relative reconstruction
efficiency over the $D^0 \to K_S^0\,\pi^+\,\pi^-$ Dalitz plot. The
considered decay chain, starting with $D^{*+}$, is simulated by means
of the {\sc evtgen} package~\cite{Lange:2001}, where the three-body
decay structure of the $D^0$ is generated without any intermediate
resonances. The generated events are passed through the detector
simulation and reconstructed as data. The simulated detector and
trigger acceptance influence the Dalitz-plot distribution in a
complicated way. To estimate the efficiency, a binned maximum
likelihood fit to the Dalitz-plot distribution of simulated decays is
performed, where a binning of $0.05\,\mathrm{GeV}^2/c^4$ in both
dimensions is used. An empiric function consisting of the sum of a
ninth-order multinomial in $(M_{K_S^0\pi^{\pm}(\mathrm{RS})}^2)^m
(M_{\pi^+\pi^-}^2)^n$, where $m+n \le 9$, and a Gaussian function,
\begin{equation}
  \begin{split}
    \epsilon & = E_0 + E_x M_{K_S^0\pi^{\pm}(\mathrm{RS})}^2 + E_y M_{\pi^+\pi^-}^2 + E_{x^2} (M_{K_S^0\pi^{\pm}(\mathrm{RS})}^2)^2 \\
    & + E_{xy} M_{K_S^0\pi^{\pm}(\mathrm{RS})}^2 M_{\pi^+\pi^-}^2 + E_{y^2} (M_{\pi^+\pi^-}^2)^2 + ... \\
    & + G(M_{\pi^+\pi^-}^2)\,,
    \label{eq:Efficiency_Func}
  \end{split}
\end{equation}
is employed. The subscripts $x$ and $y$ are abbreviations for
$M_{K_S^0\pi^{\pm}(\mathrm{RS})}^2$ and $M_{\pi^+\pi^-}^2$, respectively. The Gaussian
function $G(M_{\pi^+\pi^-}^2)$ models the efficiency enhancement at
low $M_{\pi^+\pi^-}^2$ values, which is caused by a trigger
configuration that selects track pairs with small opening angle.  The
fit projections together with the corresponding mass-squared
distributions of the three two-body combinations are shown in
Fig.~\ref{fig:EfficiencyFitProj}.

The background in the Dalitz-plot distribution receives three dominant
contributions, combinatorial background of purely random particle
combinations, misreconstructed $D^0$ candidates peaking below the
$D^{*+}$ signal, and combinations of true $D^0$ candidates with a
random pion. The first two are estimated in a sample chosen from the
$D^0$-mass upper sideband $1.92 < M(K_S^0\pi^+\pi^-) < 1.95\,\gevcc$
with the same selection requirements used in the signal region. The
combinations of true $D^0$ candidates with a random pion are directly
determined as the mistag fraction by the Dalitz-plot fit.

To estimate the contributions of the individual resonances to
the total decay rate, the fit fractions,
\begin{equation}
  \mathrm{FF}_r = \frac{\int |a_r e^{i\delta_r} \mathcal{A}_r|^2 dM_{K_S^0\pi^{\pm}(\mathrm{RS})}^2 dM_{\pi^+\pi^-}^2}{\int |\sum_j a_j e^{i\delta_j} \mathcal{A}_j|^2 dM_{K_S^0\pi^{\pm}(\mathrm{RS})}^2 dM_{\pi^+\pi^-}^2}\,,
  \label{eq:FitFrac}
\end{equation}
are calculated from the fitted amplitudes and phases. The statistical
uncertainties on the fit fractions are determined by propagating the
uncertainties on the amplitudes and phases. This is done by generating
1000 random parameter sets of amplitudes and phases according to the
full covariance matrix of the fit and taking the standard deviation of
the distribution of the 1000 calculated fit fractions.

\begin{figure}
  \includegraphics[width=8.5cm]{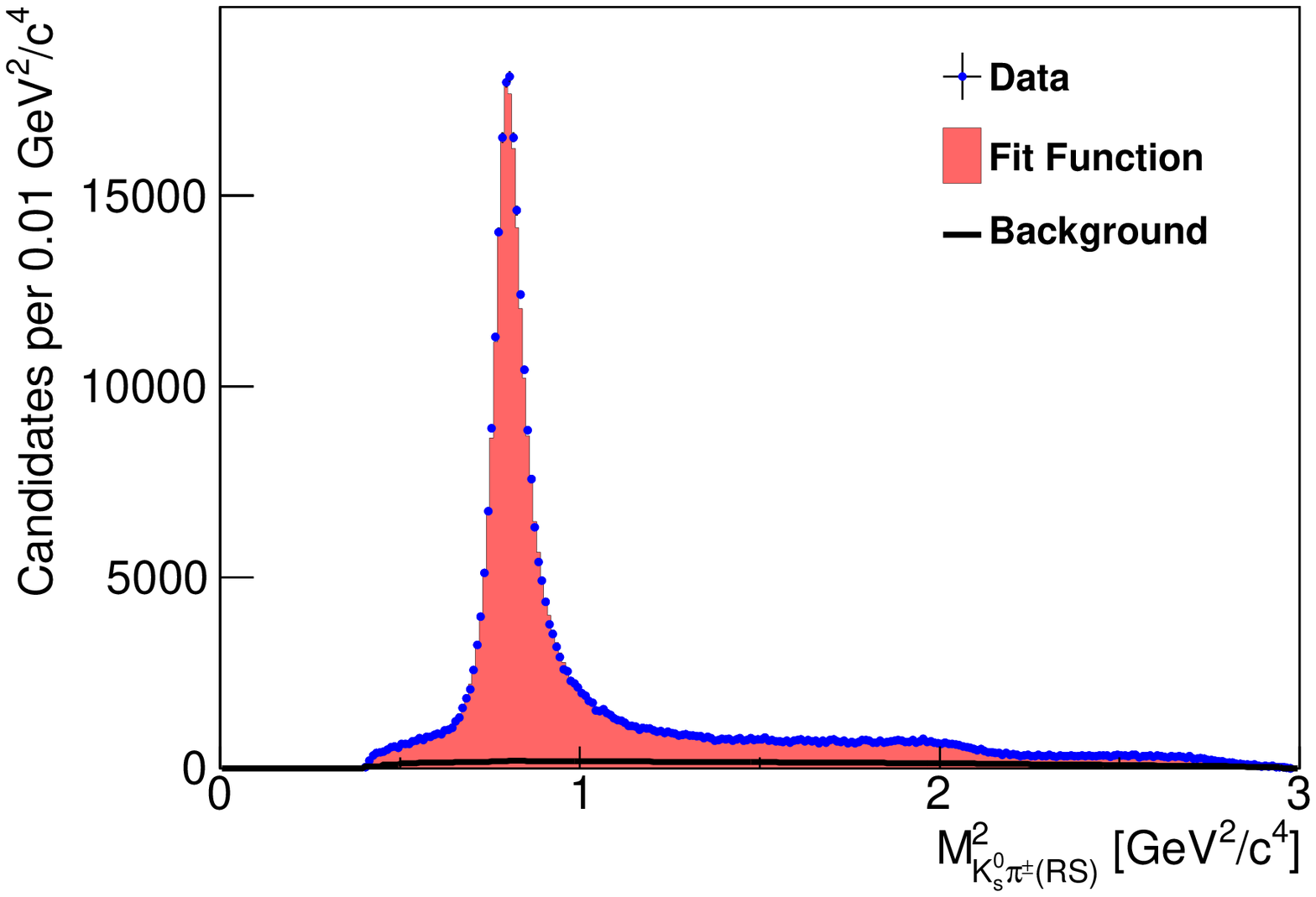}
  \includegraphics[width=8.5cm]{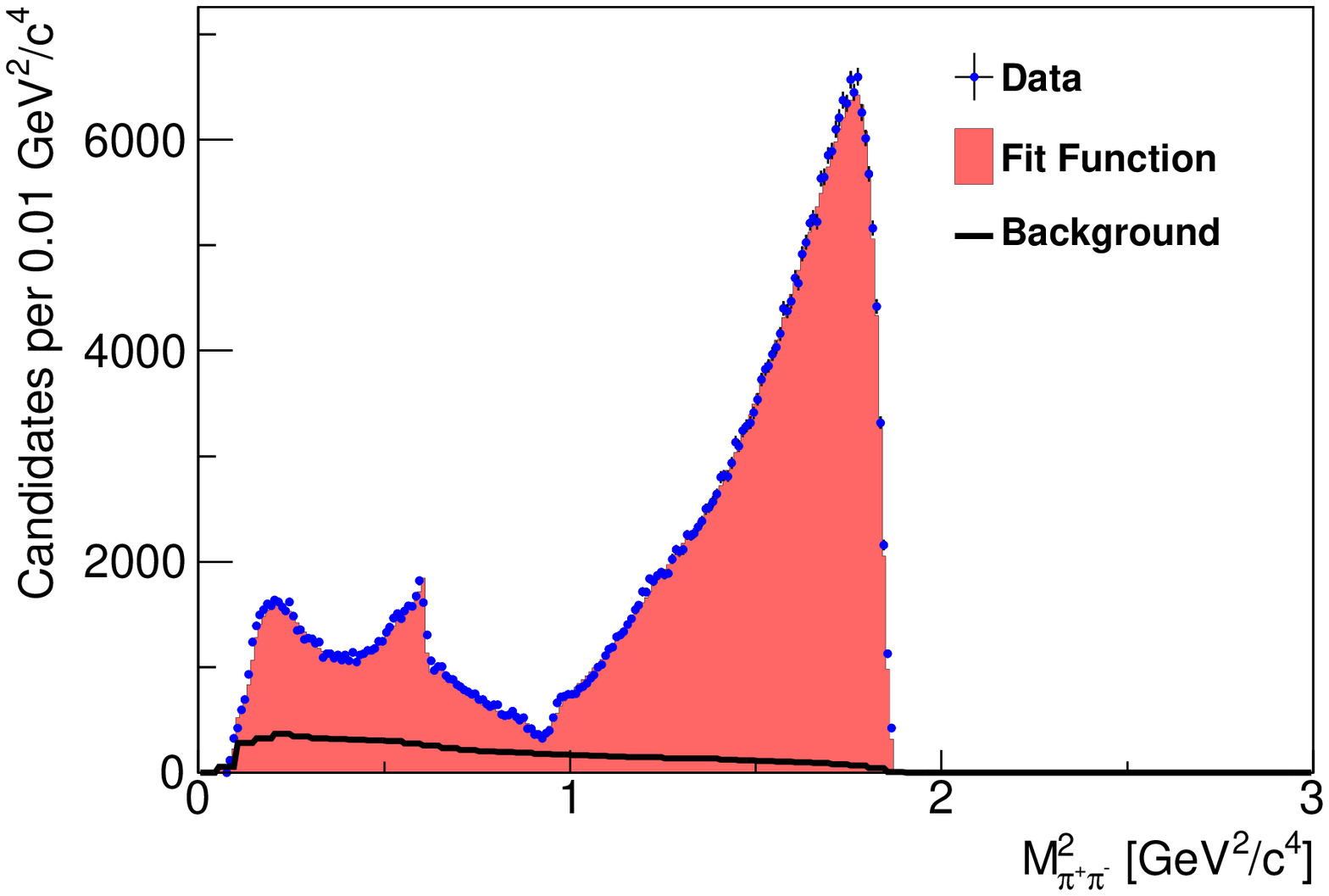}
  \includegraphics[width=8.5cm]{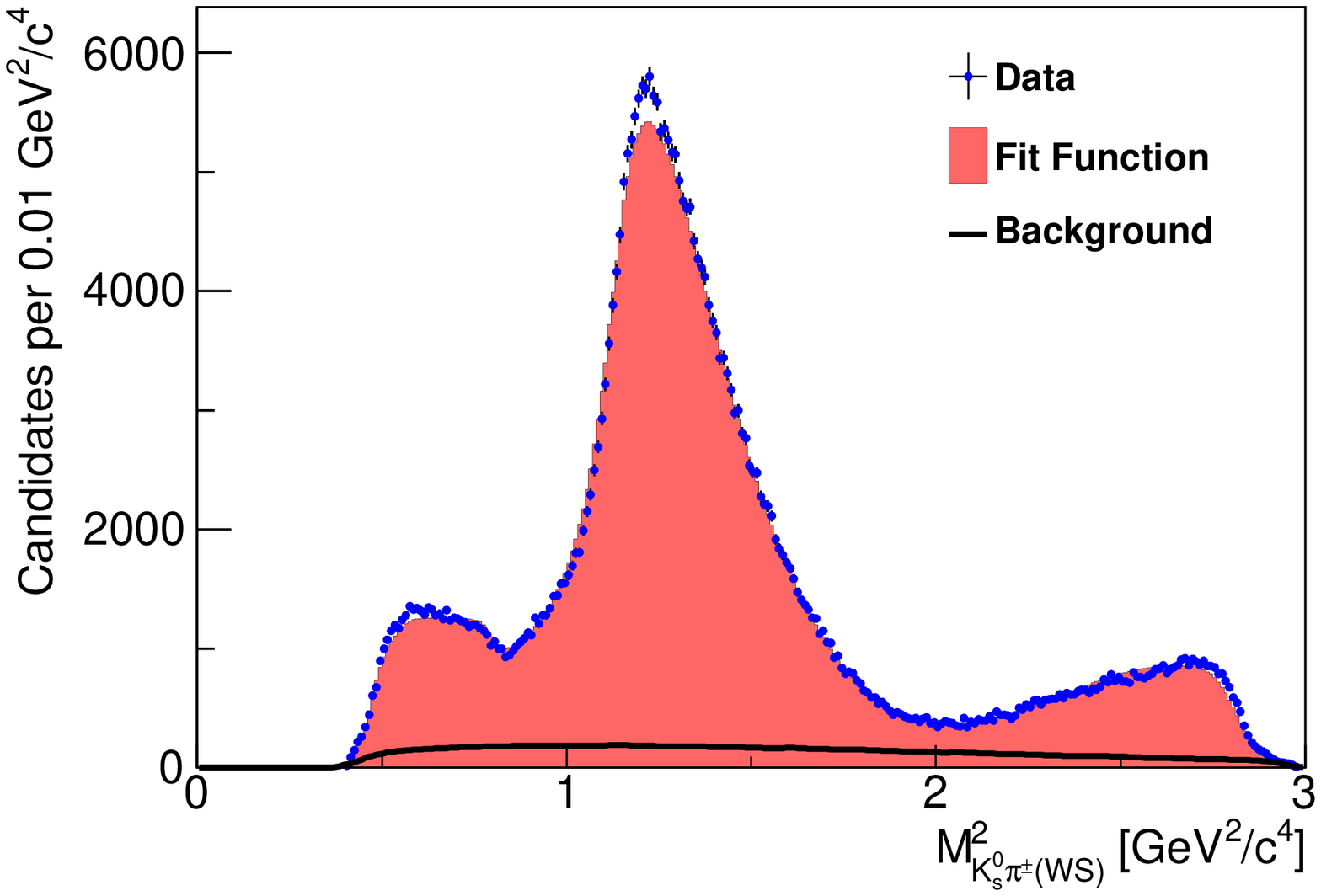}
  \caption{Projections of the Dalitz-plot fit on the individual two-body
    masses, together with the corresponding distributions in data.}
  \label{fig:DalitzFitProj}
\end{figure}

The results of the combined $D^0$ and $\bar{D}^0$ Dalitz-plot fit for
the relative amplitudes and phases of the included intermediate
resonances can be found in Table~\ref{tab:DalitzFit_Results}, together
with the corresponding fit fractions.
Table~\ref{tab:DalitzFit_MassesWidths} shows the results for the
fitted masses and widths of the $K^*(892)^{\pm}$, $f_0(600)$, and
$\sigma_2$ contributions. The values for the $K^*(892)^{\pm}$ agree
with the world-average values~\cite{PDG} within $2\,\mathrm{MeV}/c^2$.
The mistag fraction obtained from the Dalitz-plot fit is $1-T = (0.98
\pm 0.14)\%$. A reduced $\chi^2$ of $7387/5082$, calculated from the
deviations between data and fit in each bin, supports the quality of
our model. The largest discrepancy comes from the high statistics corner of the Dalitz plot populated by
the Cabibbo-favored decays with $K^{*}(892)^\pm$ resonance.
The three mass-squared projections are shown in
Fig.~\ref{fig:DalitzFitProj}. The results for the fit fractions are
consistent with the measurements from previous
experiments~\cite{CleoDalitz,Belle_gamma_old,BelleMixing,BabarMixing}.

\begin{table}
  \centering
  \caption{Combined $D^0$ and $\bar{D}^0$ Dalitz-plot-fit results for the relative amplitudes and phases of the included intermediate resonances, together with the fit fractions calculated from them. Due to interference effects between the various resonances the fit fractions are not constrained to add up exactly to 100\%.}
  \begin{tabular}{lccc}
    \hline
    \hline
    Resonance & $a$ & $\delta$ [$^\circ$] & Fit fractions [\%]\\
    \hline
    $K^*(892)^{\pm}$ & $1.911 \pm 0.012$ & $132.1 \pm 0.7$ & $61.80 \pm 0.31$\\
    $K_0^*(1430)^{\pm}$ & $2.093 \pm 0.065$ & $54.2 \pm 1.9$ & $6.25 \pm 0.25$\\
    $K_2^*(1430)^{\pm}$ & $0.986 \pm  0.034$ & $308.6 \pm 2.1$ & $1.28 \pm 0.08$\\
    $K^*(1410)^{\pm}$ & $1.092 \pm 0.069$ & $155.9 \pm 2.8$ & $1.07 \pm 0.10$\\
    $\rho(770)$ & 1 & 0 & $18.85 \pm 0.18$\\
    $\omega(782)$ & $0.038 \pm 0.002$ & $107.9 \pm 2.3$ & $0.46 \pm 0.05$\\
    $f_0(980)$ & $0.476 \pm 0.016$ & $182.8 \pm 1.3$ & $4.91 \pm 0.19$\\
    $f_2(1270)$ & $1.713 \pm 0.048$ & $329.9 \pm 1.6$ & $1.95 \pm 0.10$\\
    $f_0(1370)$ & $0.342 \pm 0.021$ & $109.3 \pm 3.1$ & $0.57 \pm 0.05$\\
    $\rho(1450)$ & $0.709 \pm 0.043$ & $8.7 \pm 2.7$ & $0.41 \pm 0.04$\\
    $f_0(600)$ & $1.134 \pm 0.041$ & $201.0 \pm 2.9$ & $7.02 \pm 0.30$\\
    $\sigma_2$ & $0.282 \pm 0.023$ & $16.2 \pm 9.0$ & $0.33 \pm 0.04$\\
    $K^*(892)^{\pm}$(DCS) & $0.137 \pm 0.007$ & $317.6 \pm 2.8$ & $0.32 \pm 0.03$\\
    $K_0^*(1430)^{\pm}$(DCS) & $0.439 \pm 0.035$ & $156.1 \pm 4.9$ & $0.28 \pm 0.04$\\
    $K_2^*(1430)^{\pm}$(DCS) & $0.291 \pm 0.034$ & $213.5 \pm 6.1$ & $0.11 \pm 0.03$\\
    Non-Resonant & $1.797 \pm 0.147$ & $94.0 \pm 5.3$ & $1.64 \pm 0.27$\\
    \hline
    Sum & & & $107.25 \pm 0.65$\\
    \hline
    \hline
  \end{tabular}
  \label{tab:DalitzFit_Results}
\end{table}

\begin{table}
  \centering
  \caption{Combined $D^0$ and $\bar{D}^0$ Dalitz-plot-fit results for the masses and widths of the $K^*(892)^{\pm}$, $f_0(600)$, and $\sigma_2$ contributions.}
  \begin{tabular}{lcc}
    \hline
    \hline
    Resonance & Mass $[\mevcc]$ & Natural width $[\mevcc]$\\
    \hline
    $K^*(892)^{\pm}$ & $893.9 \pm 0.1$ & $51.9 \pm 0.2$\\
    $f_0(600)$ & $527.3 \pm 5.2$ & $308.7 \pm 8.9$\\
    $\sigma_2$ & $1150.5 \pm 7.7$ & $138.8 \pm 7.8$\\
    \hline
    \hline
  \end{tabular}
  \label{tab:DalitzFit_MassesWidths}
\end{table}

\subsection{Measurement of \textit{CP}-violation asymmetries}
\label{sec:CPV_fit}

As described in Sec.~\ref{sec:Intro}, $D^*$ tagging is used to measure
\textit{CP}-violation effects in the Dalitz decay. Although equal
numbers of $D^0$ and $\bar{D}^0$ mesons are produced in the CDF II
detector, the efficiency for reconstructing soft pions from the $D^*$
decays causes an instrumental asymmetry between the numbers of
observed $D^0$ and $\bar{D}^0$ decays. This instrumental asymmetry is
mainly due to the tilt of COT cells described in Sec.~\ref{sec:detector},
which causes positively- and negatively-charged particles to hit the
cells at different angles. Since only relative differences between the
$D^0$ and $\bar{D}^0$ Dalitz plots are studied, an absolute efficiency
difference is expected not to bias the observed physics asymmetries.
However, an instrumental asymmetry depending on the transverse
momentum of the additional pion can lead to efficiency discrepancies
that vary over the Dalitz plot and has to be taken into account to
avoid biased results.

\begin{figure}[htb]
  \centerline{\includegraphics[width=8.5cm]{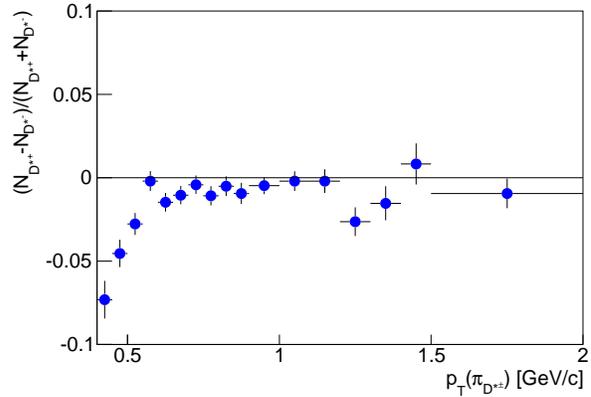}}
  \caption{Asymmetry between the numbers of reconstructed $D^{*+}$ and
    $D^{*-}$ candidates as a function of the soft pion's $p_T$.}
  \label{fig:ChargeAsym}
\end{figure}
Fig.~\ref{fig:ChargeAsym} shows the observed asymmetry,
\begin{equation}
  A = \frac{N_{D^{*+}} - N_{D^{*-}}}{N_{D^{*+}} + N_{D^{*-}}}\,,
  \label{eq:charge_asym}
\end{equation}
between the number of $D^{*+}$ ($D^0$) and $D^{*-}$ ($\bar{D}^0$)
candidates as a function of the transverse momentum of the pion from
the $D^{*\pm}$ decay. The asymmetry is larger at low
$p_T(\pi_{D^{*\pm}})$. This means that the efficiency for
reconstructing a $D^0$ or a $\bar{D}^0$ may differ over the Dalitz
plot. The effect is corrected by reweighting the $\bar{D}^0$ Dalitz
plot according to the deviations between the $p_T(\pi_{D^{*\pm}})$
distributions for positive and negative pion charges found in data.

Two different parametrization approaches to measure
\textit{CP}-violation asymmetries in a simultaneous Dalitz-plot fit to
the $D^0$ and reweighted $\bar{D}^0$ samples are applied. The first
one corresponds to an independent parametrization of the relative
amplitudes and phases in the Dalitz-plot fits of the $D^0$ and
$\bar{D}^0$ samples, respectively. Differences in the estimated
resonance parameters can then be interpreted as \textit{CP}-violation
effects. The second parametrization approach is a simultaneous fit to
the $D^0$ and $\bar{D}^0$ samples, where two additional parameters,
representing \textit{CP}-violation amplitudes and phases, are
introduced for each resonance.

\subsubsection{Independent $D^0$ and $\bar{D}^0$ parametrizations}

The fitting procedure described in Sec.~\ref{sec:comb_fit} is repeated
with separate parametrizations for the amplitudes and phases in the
$D^0$ and $\bar{D}^0$ samples. By performing a
simultaneous $D^0$ and $\bar{D}^0$ fit, common parameters are used for
the Gaussian-constrained masses and widths of the included resonances,
the non-resonant contribution, the $K^*(892)^{\pm}$, $f_0(600)$, and
$\sigma_2$ masses and widths, as well as the mistag fraction.

To quantify possible \textit{CP}-violation effects, the fit-fraction
asymmetries,
\begin{equation}
  A_{\mathrm{FF}} = \frac{\mathrm{FF}_{D^0} - \mathrm{FF}_{\bar{D}^0}}{\mathrm{FF}_{D^0} + \mathrm{FF}_{\bar{D}^0}}\,,
  \label{eq:AFF}
\end{equation}
are calculated for each intermediate resonance, where the statistical
uncertainties are determined by Gaussian uncertainties propagated from
the statistical uncertainties of the individual fit fractions.

A measure for the overall integrated \textit{CP} asymmetry is given by
\begin{equation}
  A_{CP} = \frac{\int \frac{|\mathcal{M}|^2 - |\overline{\mathcal{M}}|^2}{|\mathcal{M}|^2 + |\overline{\mathcal{M}}|^2} dM_{K_S^0\pi^{\pm}(\mathrm{RS})}^2 dM_{\pi^+\pi^-}^2}{\int dM_{K_S^0\pi^{\pm}(\mathrm{RS})}^2 dM_{\pi^+\pi^-}^2}\,,
  \label{eq:ACP}
\end{equation}
where $\mathcal{M}$ is the matrix element of Eq.~(\ref{eq:Isobar}) for
the $D^0$ decay and $\overline{\mathcal{M}}$ the one for the
$\bar{D}^0$ decay. The statistical uncertainty on $A_{CP}$ is
determined with the same procedure used for the determination of the
fit-fraction uncertainties.

\subsubsection{\textit{CP}-violation amplitudes and phases}

Following Ref.~\cite{CleoCPV}, a simultaneous fit to the $D^0$ and
$\bar{D}^0$ samples is performed, where the matrix elements for $D^0$
and $\bar{D}^0$ read
\begin{equation}
   \begin{split}
     \mathcal{M} & = a_0 e^{i\delta_0} + \sum_j a_j e^{i(\delta_j+\phi_j)}
\left(1+\frac{b_j}{a_j}\right) \mathcal{A}_j\,,\\
     \overline{\mathcal{M}} & = a_0 e^{i\delta_0} + \sum_j a_j e^{i(\delta_j-\phi_j)}
\left(1-\frac{b_j}{a_j}\right) \mathcal{A}_j\,.
   \end{split}
  \label{eq:SimDalitzFit}
\end{equation}
Compared to Eq.~(\ref{eq:Isobar}) the additional parameters $b_j$ and
$\phi_j$, representing \textit{CP}-violation amplitudes and phases,
are introduced. Again, common parameters are used for the
Gaussian-constrained masses and widths of the included resonances, the
non-resonant contribution, the $K^*(892)^{\pm}$, $f_0(600)$, and
$\sigma_2$ masses and widths, as well as the mistag fraction.

\section{Systematic uncertainties}
\label{sec:systematics}

The systematic uncertainties are categorized into experimental and
modeling uncertainties. The considered experimental sources are
efficiency asymmetries varying over the Dalitz plot, asymmetries of
the background in the $D^0$ and $\bar{D}^0$ samples, and the applied
efficiency distribution which is estimated by simulated events and
may not adequately model the composition of trigger configurations in
data. Modeling uncertainties arise from the chosen values for the
radius parameters in the Blatt-Weisskopf form factors and the limited
knowledge on the complex dynamics of the three-body decay. In this
context, the stability of the determined \textit{CP}-violation
quantities under variations of the employed Dalitz model is tested.
The contributions from the various sources to the total systematic
uncertainties can be found in
Tables~\ref{tab:D0CPV_syst_AFF}--\ref{tab:D0CPV_syst_ACP}.

\begin{table*}
  \centering
  \caption{Systematic uncertainties on the fit-fraction asymmetry $A_{\mathrm{FF}}$ for each included intermediate resonance. The contributions from the efficiency asymmetry, the background asymmetry, the fit model, the trigger efficiency, and the Blatt-Weisskopf form factors are described in Sec.~\ref{sec:systematics}. The small values for the $\rho(770)$ resonance are due to the fixing of its parameters in the fit. The total systematic uncertainties are given by adding up the various contributions in quadrature. The corresponding statistical uncertainties are listed for comparison.}
  \begin{tabular}{lccccccc}
    \hline
    \hline
    &&&&&& \multicolumn{2}{c}{Total uncertainties} \\ \cline{7-8}
    $A_{\mathrm{FF}}$ $[10^{-2}]$ & Efficiency & Background & Fit model & Trigger & Form factors & Systematic & Statistical\\
    \hline
    $K^*(892)^{\pm}$ & 0.09 & 0.13 & 0.27 & 0.09 & 0.23 & 0.40 & 0.33\\
    $K_0^*(1430)^{\pm}$ & 0.9 & 0.5 & 3.2 & 1.1 & 1.3 & 3.8 & 2.4\\
    $K_2^*(1430)^{\pm}$ & 0.8 & 0.3 & 3.9 & 0.1 & 0.6 & 4.1 & 4.0\\
    $K^*(1410)^{\pm}$ & 0.4 & 1.0 & 6.2 & 0.6 & 0.9 & 6.4 & 5.7\\
    $\rho(770)$ & 0.05 & 0.00 & 0.05 & 0.00 & 0.04 & 0.08 & 0.50\\
    $\omega(782)$ & 0.6 & 0.3 & 1.1 & 1.0 & 2.0 & 2.6 & 6.0\\
    $f_0(980)$ & 0.9 & 1.1 & 0.8 & 0.1 & 0.4 & 1.6 & 2.2\\
    $f_2(1270)$ & 1.0 & 1.4 & 1.0 & 1.5 & 1.8 & 3.0 & 3.4\\
    $f_0(1370)$ & 0.8 & 0.4 & 6.4 & 0.4 & 4.1 & 7.7 & 4.6\\
    $\rho(1450)$ & 0.6 & 0.9 & 4.7 & 6.2 & 1.9 & 8.1 & 5.2\\
    $f_0(600)$ & 0.2 & 0.2 & 2.9 & 0.2 & 2.1 & 3.6 & 2.7\\
    $\sigma_2$ & 0.9 & 1.9 & 2.5 & 0.5 & 1.9 & 3.8 & 7.6\\
    $K^*(892)^{\pm}$(DCS) & 0.3 & 0.6 & 1.3 & 0.1 & 1.5 & 2.1 & 5.7\\
    $K_0^*(1430)^{\pm}$(DCS) & 0 & 0 & 5 & 8 & 5 & 10 & 11\\
    $K_2^*(1430)^{\pm}$(DCS) & 0 & 1 & 14 & 1 & 26 & 29 & 14\\
    \hline
    \hline
  \end{tabular}
  \label{tab:D0CPV_syst_AFF}
\end{table*}

\begin{table*}
  \centering
  \caption{Systematic uncertainties on the \textit{CP}-violation amplitude $b$ for each included intermediate resonance. Further explanations can be found in the caption of Table~\ref{tab:D0CPV_syst_AFF}.}
  \begin{tabular}{lccccccc}
    \hline
    \hline
    &&&&&& \multicolumn{2}{c}{Total uncertainties} \\ \cline{7-8}
    $b$  & Efficiency & Background & Fit model & Trigger & Form factors & Systematic & Statistical\\
    \hline
    $K^*(892)^{\pm}$ & 0.011 & 0.001 & 0.000 & 0.001 & 0.002 & 0.011 & 0.004\\
    $K_0^*(1430)^{\pm}$ & 0.022 & 0.009 & 0.029 & 0.009 & 0.015 & 0.041 & 0.028\\
    $K_2^*(1430)^{\pm}$ & 0.003 & 0.004 & 0.022 & 0.003 & 0.001 & 0.023 & 0.024\\
    $K^*(1410)^{\pm}$ & 0.006 & 0.000 & 0.015 & 0.011 & 0.007 & 0.021 & 0.037\\
    $\rho(770)$ & 0.007 & 0.003 & 0.003 & 0.001 & 0.001 & 0.008 & 0.006\\
    $\omega(782)$ & 0.000 & 0.000 & 0.000 & 0.000 & 0.000 & 0.000 & 0.002\\
    $f_0(980)$ & 0.002 & 0.003 & 0.001 & 0.000 & 0.001 & 0.004 & 0.005\\
    $f_2(1270)$ & 0.003 & 0.008 & 0.008 & 0.003 & 0.005 & 0.013 & 0.037\\
    $f_0(1370)$ & 0.002 & 0.005 & 0.008 & 0.008 & 0.017 & 0.021 & 0.008\\
    $\rho(1450)$ & 0.016 & 0.023 & 0.107 & 0.071 & 0.032 & 0.135 & 0.022\\
    $f_0(600)$ & 0.011 & 0.001 & 0.018 & 0.002 & 0.013 & 0.025 & 0.017\\
    $\sigma_2$ & 0.002 & 0.002 & 0.002 & 0.001 & 0.001 & 0.004 & 0.012\\
    $K^*(892)^{\pm}$(DCS) & 0.001 & 0.001 & 0.001 & 0.000 & 0.001 & 0.002 & 0.005\\
    $K_0^*(1430)^{\pm}$(DCS) & 0.003 & 0.005 & 0.030 & 0.008 & 0.015 & 0.035 & 0.024\\
    $K_2^*(1430)^{\pm}$(DCS) & 0.005 & 0.001 & 0.006 & 0.002 & 0.015 & 0.017 & 0.029\\
    \hline
    \hline
  \end{tabular}
  \label{tab:D0CPV_syst_b}
\end{table*}

\begin{table*}
  \centering
  \caption{Systematic uncertainties on the \textit{CP}-violation phase $\phi$ for each included intermediate resonance. Further explanations can be found in the caption of Table~\ref{tab:D0CPV_syst_AFF}.}
  \begin{tabular}{lccccccc}
    \hline
    \hline
    &&&&&& \multicolumn{2}{c}{Total uncertainties} \\ \cline{7-8}
    $\phi$ [$^\circ$] & Efficiency & Background & Fit model & Trigger & Form factors & Systematic & Statistical\\
    \hline
    $K^*(892)^{\pm}$ & 0.0 & 0.1 & 1.0 & 0.2 & 0.8 & 1.3 & 1.4\\
    $K_0^*(1430)^{\pm}$ & 0.0 & 0.0 & 2.1 & 0.2 & 0.7 & 2.2 & 1.7\\
    $K_2^*(1430)^{\pm}$ & 0.1 & 0.2 & 1.0 & 0.3 & 0.4 & 1.1 & 1.8\\
    $K^*(1410)^{\pm}$ & 0.1 & 0.0 & 0.8 & 1.3 & 1.6 & 2.2 & 1.9\\
    $\rho(770)$ & 0.1 & 0.3 & 1.1 & 0.2 & 0.8 & 1.4 & 1.5\\
    $\omega(782)$ & 0.3 & 0.6 & 0.9 & 0.1 & 0.8 & 1.4 & 2.2\\
    $f_0(980)$ & 0.2 & 0.0 & 0.8 & 0.0 & 0.7 & 1.1 & 1.3\\
    $f_2(1270)$ & 0.2 & 0.2 & 1.3 & 0.7 & 1.4 & 2.1 & 1.9\\
    $f_0(1370)$ & 1.0 & 1.2 & 1.9 & 1.0 & 0.9 & 2.8 & 1.7\\
    $\rho(1450)$ & 0.0 & 0.6 & 2.0 & 3.1 & 1.2 & 3.9 & 1.7\\
    $f_0(600)$ & 0.1 & 0.2 & 1.3 & 0.2 & 0.3 & 1.4 & 1.5\\
    $\sigma_2$ & 0.2 & 0.3 & 0.7 & 0.4 & 0.7 & 1.1 & 2.9\\
    $K^*(892)^{\pm}$(DCS) & 0.1 & 0.5 & 1.1 & 0.1 & 0.1 & 1.2 & 2.3\\
    $K_0^*(1430)^{\pm}$(DCS) & 0.0 & 0.3 & 3.5 & 1.0 & 1.3 & 3.9 & 4.0\\
    $K_2^*(1430)^{\pm}$(DCS) & 0.6 & 0.5 & 2.3 & 0.5 & 1.7 & 3.0 & 5.3\\
    \hline
    \hline
  \end{tabular}
  \label{tab:D0CPV_syst_phi}
\end{table*}

\begin{table*}
  \centering
  \caption{Systematic uncertainties on the overall integrated \textit{CP} asymmetry. Further explanations can be found in the caption of Table~\ref{tab:D0CPV_syst_AFF}.}
  \begin{tabular}{lc} \hline\hline
    Effect  &  Uncertainty on $A_{CP}$ $[10^{-2}]$ \\ \hline
   Efficiency & 0.36   \\
   Background & 0.09   \\
   Fit model &  0.37   \\
   Trigger  &   0.05   \\
   Form factors & 0.10 \\ \hline
   Total systematic & 0.54 \\ \hline
   Statistical  & 0.57 \\ \hline\hline
  \end{tabular}
  \label{tab:D0CPV_syst_ACP}
\end{table*}

\subsection{Efficiency asymmetry}
\label{sec:Eff.}

The reweighting procedure of the $\bar{D}^0$ Dalitz plot according to
the deviations between the $p_T(\pi_{D^{*\pm}})$ distributions for
positively and negatively charged pions may not fully correct for
residual small asymmetries between the $D^0$ and $\bar{D}^0$
efficiency distributions. To estimate the size of a systematic effect
originating from such an asymmetry, the Dalitz-plot fits are repeated
without reweighting the $\bar{D}^0$ Dalitz plot. The scale of
systematic uncertainties are estimated as the differences between the
resulting values and the ones from the default fits.

\subsection{Background asymmetry}
\label{sec:Bkg.}

To investigate a possible systematic effect originating from different
Dalitz-plot distributions of the background in $D^0$ and $\bar{D}^0$
data, the Dalitz-plot fits are repeated with two independent
background samples distinguished by the charge of the slow pion in the
$D^{*\pm}$ decay. The systematic uncertainties are calculated as
differences between the resulting values and the ones from the default
fits.

\subsection{Fit model}
\label{sec:Model}

The systematic uncertainties originating from the specific model used
for the Dalitz-plot fit are estimated by repeating the fits when one
of the resonances $K^*(1410)^{\pm}$, $f_0(1370)$, $\sigma_2$,
$K_2^*(1430)^{\pm}$(DCS), or the non-resonant contribution is excluded from the
model. 
These are the least significant contributions, or in the case of $\sigma_2$, because of the controversy over the existence of such a state.
For each case, the
resulting \textit{CP}-violation quantities are compared to the values
from the default fits and the largest deviations are used as modeling
systematic uncertainties.

\subsection{Efficiency model}
\label{sec:Trigger}

As described in Sec.~\ref{sec:detector}, different trigger
configurations with thresholds of 4, 5.5, and 6.5\,\gevc on the scalar
sum of the transverse momenta of the two trigger tracks are applied
for the online event selection. The simulated events, used to
determine the reconstruction efficiency across the Dalitz plot, are
required to pass the different trigger configurations in the same
proportions as the actual data. Any mismodeling of the efficiency is
expected to cancel between $D^0$ and $\bar{D}^0$ candidates, to first
order. To estimate residual higher-order effects, an efficiency
determined with simulated events satisfying the 6.5\,\gevc trigger
threshold is used, and the difference with the default efficiency is
taken as a systematic uncertainty.

\subsection{Blatt-Weisskopf form factors}
\label{sec:FF}

In the default Dalitz-plot fits, the chosen values for the radius
parameter $R$ in the Blatt-Weisskopf form factors are $R =
5\,(\mathrm{GeV}/c)^{-1}$ for the $D^0$ and $R =
1.5\,(\mathrm{GeV}/c)^{-1}$ for all intermediate
resonances~\cite{dalitz_method}. To estimate the systematic
uncertainties originating from deviations from these values, the
Dalitz-plot fits are repeated with both values for the radius
parameter halved and doubled. The systematic uncertainties are then
calculated as the largest differences between the resulting values and
the ones from the default fits.

\subsection{Further checks}

To assess the robustness of the fitting procedure, the following
checks were done, each leading to results that agree with the default
fits within the expected variations.

To test the effects of the Dalitz-plot binning the fits are repeated
when varying the bin widths from $0.025\,\mathrm{GeV}^2/c^4$ to
$0.03\,\mathrm{GeV}^2/c^4$ and $0.05\,\mathrm{GeV}^2/c^4$.

Discrepancies between data and the Dalitz-plot fit are visible in
Fig.~\ref{fig:DalitzFitProj}, in particular at the peak of the
$K^*(892)^{\pm}$ signal. To verify the results, the fits are repeated when
excluding the Dalitz-plot regions with the largest discrepancies
between the fit values and data.

Instead of using Gaussian constraints, the masses and decay widths of
the included resonances are fixed to the values in
Ref.~\cite{f0masses} for $f_0(980)$ and $f_0(1370)$, and to the world
average values~\cite{PDG} for the others, except for $K^*(892)^{\pm}$,
$f_0(600)$, and $\sigma_2$ which are still unconstrained parameters in
the fit.

The Dalitz-plot fits are repeated with an alternative background
model, where the combinatorial background and the combinations of true
$D^0$ candidates with a random pion are modeled by $D^{*+}$ sidebands
and the distribution of misreconstructed $D^0$ candidates is modeled
by means of an inclusive charm simulated dataset. The ratio of
combinatorial background and combinations of true $D^0$ candidates
with a random pion is assumed to be independent of the $D^{*+}$ mass
difference distribution. Since this assumption is not completely
valid, especially close to the kinematic threshold, the method is only
chosen as check of the default background model described in
Sec.~\ref{sec:comb_fit}.

\section{Results}
\label{sec:results}

All \textit{CP}-violation quantities are found to be consistent with
zero. The results for the \textit{CP}-violation amplitudes and phases,
defined in Eq.~(\ref{eq:SimDalitzFit}) and obtained from the
simultaneous fit to the $D^0$ and $\bar{D}^0$ Dalitz plots, are
displayed in Table~\ref{tab:D0CPV_Results2}. The fit-fraction
asymmetries for the intermediate resonances, defined in
Eq.~(\ref{eq:AFF}), are listed in Table~\ref{tab:D0CPV_Results1}. The
overall time-integrated \textit{CP} asymmetry, defined in
Eq.~(\ref{eq:ACP}), is determined to be
\begin{equation}
  A_{CP} = (-0.05 \pm 0.57 (\mathrm{stat}) \pm 0.54 (\mathrm{syst}))\%\,.
\end{equation}
This value includes the contribution from time-integrated \textit{CP}
violation in the mixing of the involved $K^0$ mesons. We determine
this contribution with the method described in
Ref.~\cite{Grossman:2011zk} to be $-0.07\%$, much smaller than the
$A_{CP}$ measurement uncertainty.

\begin{table}
  \centering
  \caption{Results of the simultaneous $D^0$-$\bar{D}^0$ Dalitz-plot fit for the \textit{CP}-violation amplitudes, $b$, and phases, $\phi$. The first uncertainties are statistical and the second systematic.}
  \begin{tabular}{lcccc}
    \hline
    \hline
    Resonance & $b$ & $\phi$ [$^\circ$]\\
    \hline
    $K^*(892)^{\pm}$ & $+0.004 \pm 0.004 \pm 0.011$ & $-0.8 \pm 1.4 \pm 1.3$\\
    $K_0^*(1430)^{\pm}$ & $+0.044 \pm 0.028 \pm 0.041$ & $-1.8 \pm 1.7 \pm 2.2$\\
    $K_2^*(1430)^{\pm}$ & $+0.018 \pm 0.024 \pm 0.023$ & $-1.1 \pm 1.8 \pm 1.1$\\
    $K^*(1410)^{\pm}$ & $-0.010 \pm 0.037 \pm 0.021$ & $-1.6 \pm 1.9 \pm 2.2$\\
    $\rho(770)$ & $-0.003 \pm 0.006 \pm 0.008$ & $-0.5 \pm 1.5 \pm 1.4$\\
    $\omega(782)$ & $-0.003 \pm 0.002 \pm 0.000$ & $-1.8 \pm 2.2 \pm 1.4$\\
    $f_0(980)$ & $-0.001 \pm 0.005 \pm 0.004$ & $-0.1 \pm 1.3 \pm 1.1$\\
    $f_2(1270)$ & $-0.035 \pm 0.037 \pm 0.013$ & $-2.0 \pm 1.9 \pm 2.1$\\
    $f_0(1370)$ & $-0.002 \pm 0.008 \pm 0.021$ & $-0.1 \pm 1.7 \pm 2.8$\\
    $\rho(1450)$ & $-0.016 \pm 0.022 \pm 0.135$ & $-1.7 \pm 1.7 \pm 3.9$\\
    $f_0(600)$ & $-0.012 \pm 0.017 \pm 0.025$ & $-0.3 \pm 1.5 \pm 1.4$\\
    $\sigma_2$ & $-0.011 \pm 0.012 \pm 0.004$ & $-0.2 \pm 2.9 \pm 1.1$\\
    $K^*(892)^{\pm}$(DCS) & $+0.001 \pm 0.005 \pm 0.002$ & $-3.8 \pm 2.3 \pm 1.2$\\
    $K_0^*(1430)^{\pm}$(DCS) & $+0.022 \pm 0.024 \pm 0.035$ & $-3.3 \pm 4.0 \pm 3.9$\\
    $K_2^*(1430)^{\pm}$(DCS) & $-0.018 \pm 0.029 \pm 0.017$ & $+4.2 \pm 5.3 \pm 3.0$\\
    \hline
    \hline
  \end{tabular}
  \label{tab:D0CPV_Results2}
\end{table}

\begin{table}
  \centering
  \caption{Fit-fraction asymmetries, $A_{\mathrm{FF}}$, for the included intermediate resonances. The first uncertainties are statistical and the second systematic.}
  \begin{tabular}{lc}
    \hline
    \hline
    Resonance & $A_{\mathrm{FF}}$ [\%]\\
    \hline
    $K^*(892)^{\pm}$ & $+0.36 \pm 0.33 \pm 0.40$\\
    $K_0^*(1430)^{\pm}$ & $+4.0 \pm 2.4 \pm 3.8$\\
    $K_2^*(1430)^{\pm}$ & $+2.9 \pm 4.0 \pm 4.1$\\
    $K^*(1410)^{\pm}$ & $-2.3 \pm 5.7 \pm 6.4$\\
    $\rho(770)$ & $-0.05 \pm 0.50 \pm 0.08$\\
    $\omega(782)$ & $-12.6 \pm 6.0 \pm 2.6$\\
    $f_0(980)$ & $-0.4 \pm 2.2 \pm 1.6$\\
    $f_2(1270)$ & $-4.0 \pm 3.4 \pm 3.0$\\
    $f_0(1370)$ & $-0.5 \pm 4.6 \pm 7.7$\\
    $\rho(1450)$ & $-4.1 \pm 5.2 \pm 8.1$\\
    $f_0(600)$ & $-2.7 \pm 2.7 \pm 3.6$\\
    $\sigma_2$ & $-6.8 \pm 7.6 \pm 3.8$\\
    $K^*(892)^{\pm}$(DCS) & $+1.0 \pm 5.7 \pm 2.1$\\
    $K_0^*(1430)^{\pm}$(DCS) & $+12 \pm 11 \pm 10$\\
    $K_2^*(1430)^{\pm}$(DCS) & $-10 \pm 14 \pm 29$\\
    \hline
    \hline
  \end{tabular}
  \label{tab:D0CPV_Results1}
\end{table}

\subsection{Indirect \textit{CP} violation}

Following the procedure described in Ref.~\cite{D0hh}, it is possible
to disentangle indirect from direct \textit{CP}-violation effects by means of
the $D^0$ decay time distribution. The direct and indirect \textit{CP}
asymmetries are related to the time-integrated asymmetry by
\begin{equation}
  A_{CP} = A_{CP}^{\mathrm{dir}} + \frac{\langle t \rangle}{\tau} A_{CP}^{\mathrm{ind}}\,,
\end{equation}
where $\tau$ is the mean $D^0$ lifetime. The mean observed decay
time $\langle t \rangle$ is determined from the background subtracted
$D^0$ decay time distribution. We correct for the fraction of
nonprompt events that is estimated from the $D^{*+}$ impact parameter
significance distribution, and obtain a mean observed decay time of $\langle
t \rangle = (2.28 \pm 0.03) \tau(D^0)$. To compare with the recent CDF
measurement of \textit{CP}-violation asymmetries in $D^0 \to
\pi^+\pi^-$ and $D^0 \to K^+K^-$ decays, $A_{CP}^\mathrm{ind}(D^0 \to
h^+h^-) = (-0.01 \pm 0.06 (\mathrm{stat}) \pm 0.04
(\mathrm{syst}))\%$~\cite{D0hh}, we determine the indirect \textit{CP} 
asymmetry for the case of no direct \textit{CP} violation to be
\begin{equation}
  A_{CP}^\mathrm{ind} = (-0.02 \pm 0.25 (\mathrm{stat}) \pm 0.24 (\mathrm{syst}))\%\,.
\end{equation}

\subsection{Individual \textit{CP}-violation asymmetries}

The CLEO experiment also quotes \textit{CP}-violation quantities
called interference fractions, $\mathrm{IF}$, and individual
\textit{CP} asymmetries, $A_{CP}$, in each
subresonance~\cite{CleoCPV}. These are defined as $A_{CP_j} =
\frac{\mathrm{IF}_j}{\mathrm{FF}_j}$, where
\begin{widetext}
\begin{equation}
  \mathrm{IF}_j = \frac{|\int \sum_k (2 a_k e^{i\delta_k} \sin(\phi_k + \phi_j) \mathcal{A}_k) b_j \mathcal{A}_j dM_{K_S^0\pi^{\pm}(\mathrm{RS})}^2 dM_{\pi^+\pi^-}^2|}{\left(\int |\mathcal{M}|^2 dM_{K_S^0\pi^{\pm}(\mathrm{RS})}^2 dM_{\pi^+\pi^-}^2 + \int |\overline{\mathcal{M}}|^2 dM_{K_S^0\pi^{\pm}(\mathrm{RS})}^2 dM_{\pi^+\pi^-}^2\right)}\,.
\end{equation}
\end{widetext}
Since these values are positive by construction, only upper limits are
given. The calculation is performed with the same method used for the
determination of the fit fractions, where the 90\% and 95\% quantiles
of resulting distributions are used as the corresponding C.L. upper
limits. To account for systematic uncertainties for each resonance,
the largest values of all fits with the different systematic
variations are taken. The resulting 90\% and 95\% C.L. on the
individual \textit{CP} asymmetries are listed in
Table~\ref{tab:ACP_Limits}.

\begin{table}
  \centering
  \caption{Upper limits for individual \textit{CP}-violation asymmetries.}
  \begin{tabular}{lcc}
    \hline
    \hline
    Resonance & $A_{CP}$ [\%] (90\% C.L.) & $A_{CP}$ [\%] (95\% C.L.)\\
    \hline
    $K^*(892)^{\pm}$ & $0.014$ & $0.018$\\
    $K_0^*(1430)^{\pm}$ & $0.80$ & $1.2$\\
    $K_2^*(1430)^{\pm}$ & $0.45$ & $0.62$\\
    $K^*(1410)^{\pm}$ & $6.6$ & $8.4$\\
    $\rho(770)$ & $0.038$ & $0.051$\\
    $\omega(782)$ & $0.51$ & $0.66$\\
    $f_0(980)$ & $0.13$ & $0.17$\\
    $f_2(1270)$ & $1.6$ & $2.1$\\
    $f_0(1370)$ & $25$ & $37$\\
    $\rho^0(1450)$ & $6.5$ & $8.2$\\
    $f_0(600)$ & $0.17$ & $0.24$\\
    $\sigma_2$ & $3.1$ & $3.9$\\
    $K^*(892)^{\pm}$(DCS) & $1.7$ & $2.3$\\
    $K_0^*(1430)^{\pm}$(DCS) & $22$ & $28$\\
    $K_2^*(1430)^{\pm}$(DCS) & $12$ & $16$\\
    \hline
    \hline
  \end{tabular}
  \label{tab:ACP_Limits}
\end{table}

\section{Model-independent approach}
\label{sec:CPV_Binned}

Following Ref.~\cite{Bigi}, a model-independent search for \textit{CP}
violation in the Dalitz-plot distribution of the decay $D^0 \to
K_S^0\,\pi^+\,\pi^-$ is performed by comparing the binned 
Dalitz plots for $D^0$ and $\bar{D}^0$ meson decays. No assumptions
about the resonant substructure of the decay are used. The approach
serves as a complementary verification of the results from the
Dalitz-plot fits described in the previous Sections. However, this
method only detects the presence of a significant
\textit{CP}-violation effect, without allowing a determination of the
size of the asymmetries.

The signed significance of the asymmetry between the numbers of $D^0$
and $\bar{D}^0$ candidates,
$(N_{D^0}-N_{\bar{D}^0})/\sqrt{N_{D^0}+N_{\bar{D}^0}}$, is calculated
for each bin and studied as a function of the squared $K_S^0\pi^{\pm}$
and $\pi^+\pi^-$ masses. In this calculation the number of $\bar{D}^0$ events is normalized to the one
of $D^0$.
Possible \textit{CP}-violation asymmetries
would appear as clusters of same-sign discrepancies. The sum of the
squares of the significance asymmetries in each bin is expected to
follow a $\chi^2$ distribution. The $p$-value can be calculated
considering the number of degrees of freedom equal to the number of
Dalitz-plot bins minus one (for the normalization). Furthermore, one
expects a Gaussian distribution with 
mean equal to 0 and width of 1 for the
histogram of the asymmetry significance distribution in case of
vanishing \textit{CP} violation.

The method is verified in simulation and then applied to data. 
As we test relative differences between $D^0$ and $\bar{D}^0$ at different places in the Dalitz plot, we
normalize $D^0$ and $\bar{D}^0$ to the same area. With this approach, all asymmetries that are
uniformly distributed over the Dalitz plot completely cancel.
However, an efficiency asymmetry
varying over the Dalitz plot may mimic \textit{CP} violation. As
described in Sec.~\ref{sec:CPV_fit}, this problem is also relevant for
the Dalitz-plot fits, and the reweighting procedure used there is
applied here as well.

\begin{figure}[tbh]
  \includegraphics[width=8.5cm]{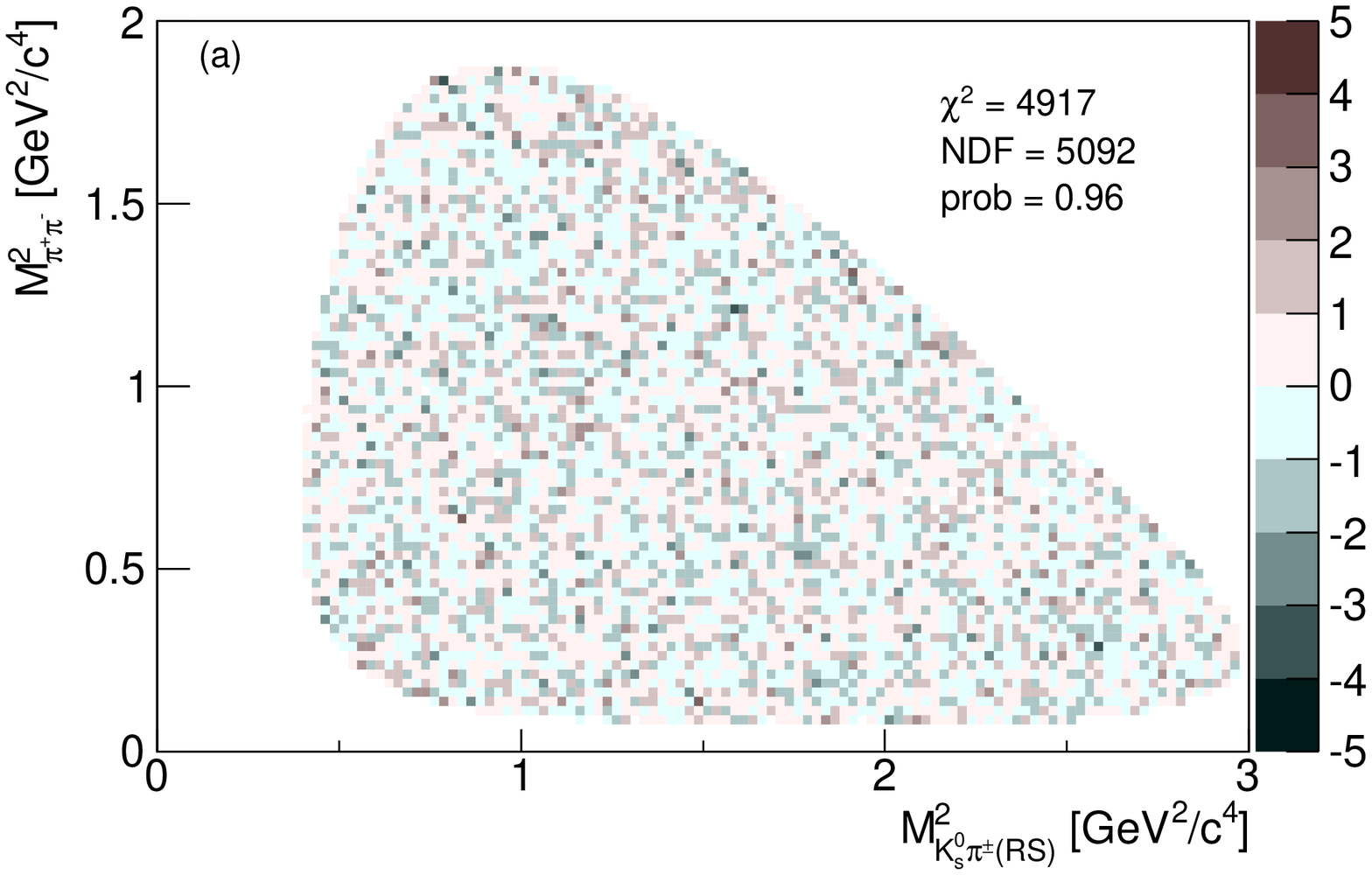}
  \includegraphics[width=8.5cm]{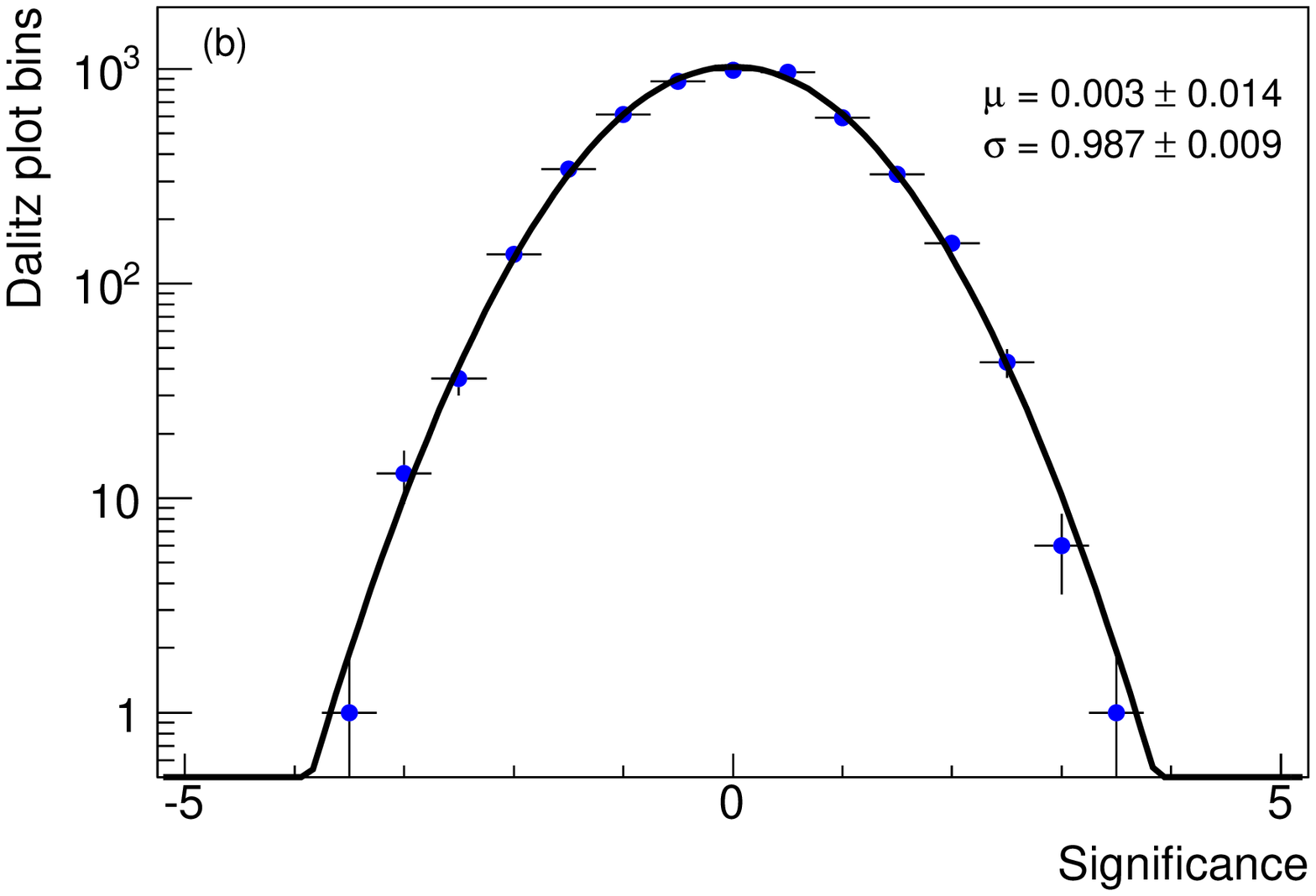}
  \caption{Distribution of (a) the asymmetry significance as a function of
    the squared $K_S^0\pi^{\pm}$ and $\pi^+\pi^-$ masses and (b) as a histogram.}
  \label{fig:Miranda}
\end{figure}
The resulting Dalitz-plot distribution of the asymmetry significance
between the numbers of $D^0$ and $\bar{D}^0$ candidates, with bin
widths of $0.025\,\mathrm{GeV}^2/c^4$ in both dimensions, and the
corresponding histogram are shown in Fig.~\ref{fig:Miranda}. The
parameters obtained from the fit, $\mu=0.003 \pm 0.014$ and
$\sigma=0.987 \pm 0.009$, are consistent with a Gaussian distribution
centered at zero with unit variance, and the $p$-value calculated from
the asymmetry significance distribution is $p=0.96$. The
model-independent approach confirms that no \textit{CP} violation is
observed between the $D^0$ and $\bar{D}^0$ decay amplitudes into the
$K_S^0\pi^+\pi^-$ final state.

\section{Conclusion}
\label{sec:conclusion}

A Dalitz-amplitude analysis is employed to study the resonant
substructure of the $D^0 \to K_S^0\,\pi^+\,\pi^-$ three-body decay. In
performing a full Dalitz-plot fit, the relative amplitudes, phases,
and fit fractions of the various intermediate resonances are
determined. The results are compatible and comparable in precision to
the measurements from previous
experiments~\cite{CleoDalitz,Belle_gamma_old,BelleMixing,BabarMixing}.

In simultaneous fits to the $D^0$ and $\bar{D}^0$ Dalitz plots, a
\textit{CP}-violation fit fraction, amplitude, and phase are
determined for each included intermediate resonance. None of these is
significantly different from zero. This also holds for the overall
integrated \textit{CP} asymmetry, $A_{CP} = (-0.05 \pm 0.57
(\mathrm{stat}) \pm 0.54 (\mathrm{syst}))\%$. A complementary
model-independent search for localized \textit{CP}-violation
differences in relative Dalitz-plot densities between the binned $D^0$
and $\bar{D}^0$ distributions yields a result consistent with zero,
too. In conclusion, the most precise values for the overall integrated
\textit{CP} asymmetry as well as the \textit{CP}-violation fit
fractions, amplitudes, and phases are reported; no indications for any
\textit{CP}-violation effects in $D^0 \to K_S^0\,\pi^+\,\pi^-$ decays
are found, in agreement with the standard model.

\begin{acknowledgments}
  We thank the Fermilab staff and the technical staffs of the
  participating institutions for their vital contributions. This work
  was supported by the U.S. Department of Energy and National Science
  Foundation; the Italian Istituto Nazionale di Fisica Nucleare; the
  Ministry of Education, Culture, Sports, Science and Technology of
  Japan; the Natural Sciences and Engineering Research Council of
  Canada; the National Science Council of the Republic of China; the
  Swiss National Science Foundation; the A.P. Sloan Foundation; the
  Bundesministerium f\"ur Bildung und Forschung, Germany; the Korean
  World Class University Program, the National Research Foundation of
  Korea; the Science and Technology Facilities Council and the Royal
  Society, UK; the Russian Foundation for Basic Research; the
  Ministerio de Ciencia e Innovaci\'{o}n, and Programa
  Consolider-Ingenio 2010, Spain; the Slovak R\&D Agency; the Academy
  of Finland; and the Australian Research Council (ARC).
\end{acknowledgments}

%

\end{document}